\pdfoutput=1
\documentclass[12pt,a4paper]{article}
\usepackage{ifthen} % for conditional statements
\newboolean{pdflatex}
\setboolean{pdflatex}{true} % False for eps figures 

\newboolean{articletitles}
\setboolean{articletitles}{true} % False removes titles in references

\newboolean{uprightparticles}
\setboolean{uprightparticles}{true} %True for upright particle symbols

\def\paperauthors{LHCb collaboration} % Leave as is for PAPER, CONF and FIGURE
\def\paperasciititle{Observation of muonic Dalitz decays of chib mesons and precise spectroscopy 
of hidden-beauty states} % Set ASCII title here !! MAKE sure it's only ASCII characters !! 
\def\papertitle{Observation of muonic Dalitz decays of $\Pchi_\bquark$ mesons and 
precise spectroscopy of hidden-beauty states} % Latex formatted title
\def\paperkeywords{{High Energy Physics}, {LHCb}} % Comma separated list
\def\papercopyright{\the\year\ CERN for the benefit of the LHCb collaboration} % new since 9/Apr/2018
\def\paperlicence{CC BY 4.0 licence}
\def\paperlicenceurl{https://creativecommons.org/licenses/by/4.0/}

\newif\ifEnableSectionTOCLinks
\EnableSectionTOCLinksfalse % deactivated

\usepackage{rotating}
\usepackage{amsmath}
\usepackage{mathrsfs} 
\usepackage{graphpap} 
\usepackage{lscape} 
\usepackage{multirow} 
\usepackage{tikz}
\usetikzlibrary{patterns.meta}
\usepackage{xcolor}
\usepackage{transparent}
\usepackage{wasysym}
\usepackage{dashrule}
\usepackage{scalerel}
\DeclareMathOperator*{\bigplus}{\scalerel*{+}{\sum}}

\makeatletter
\g@addto@macro\bfseries{\boldmath}
\makeatother

\usepackage[top=1in, bottom=1.25in, left=1in, right=1in]{geometry}

\usepackage{microtype}
\usepackage{lineno}  % for line numbering during review
\usepackage{xspace} % To avoid problems with missing or double spaces after
\usepackage{caption} %these three command get the figure and table captions automatically small

\usepackage{graphicx}  % to include figures (can also use other packages)
\usepackage{color}
\usepackage{colortbl}
\graphicspath{{./figs/}} % Make Latex search fig subdir for figures
\usepackage{amsmath} % Adds a large collection of math symbols
\usepackage{amssymb}
\usepackage{amsfonts}
\usepackage{upgreek} % Adds in support for greek letters in roman typeset

\newcommand*\patchAmsMathEnvironmentForLineno[1]{%
\expandafter\let\csname old#1\expandafter\endcsname\csname #1\endcsname
\expandafter\let\csname oldend#1\expandafter\endcsname\csname
end#1\endcsname
 \renewenvironment{#1}%
   {\linenomath\csname old#1\endcsname}%
   {\csname oldend#1\endcsname\endlinenomath}%
}
\newcommand*\patchBothAmsMathEnvironmentsForLineno[1]{%
  \patchAmsMathEnvironmentForLineno{#1}%
  \patchAmsMathEnvironmentForLineno{#1*}%
}
\AtBeginDocument{%
\patchBothAmsMathEnvironmentsForLineno{equation}%
\patchBothAmsMathEnvironmentsForLineno{align}%
\patchBothAmsMathEnvironmentsForLineno{flalign}%
\patchBothAmsMathEnvironmentsForLineno{alignat}%
\patchBothAmsMathEnvironmentsForLineno{gather}%
\patchBothAmsMathEnvironmentsForLineno{multline}%
\patchBothAmsMathEnvironmentsForLineno{eqnarray}%
}

\usepackage[pdftex,
            pdfauthor={\paperauthors},
            pdftitle={\paperasciititle},
            pdfkeywords={\paperkeywords},
			]{hyperref}

\usepackage{hyperxmp}
			
\usepackage[colorinlistoftodos,textsize=scriptsize]{todonotes}

\usepackage[bottom,flushmargin,hang,multiple]{footmisc}

\usepackage[all]{hypcap} % Internal hyperlinks to floats.

\usepackage{xspace} 
\usepackage{upgreek}

\def\lhcb   {\mbox{LHCb}\xspace}

\def\MagUp {\mbox{\em Mag\kern -0.05em Up}\xspace}

\ifthenelse{\boolean{uprightparticles}}%
{
 
 \def\Pgamma      {\ensuremath{\upgamma}\xspace}

 \def\Pmu         {\ensuremath{\upmu}\xspace}

 \def\Ppi         {\ensuremath{\uppi}\xspace}

 \def\Pchi        {\ensuremath{\upchi}\xspace}                 
 \def\Ppsi        {\ensuremath{\uppsi}\xspace}

 \def\PDelta      {\ensuremath{\Delta}\xspace}                 
 \def\PXi         {\ensuremath{\Xi}\xspace}                 
 \def\PLambda     {\ensuremath{\Lambda}\xspace}                 
 \def\PSigma      {\ensuremath{\Sigma}\xspace}                 
 \def\POmega      {\ensuremath{\Omega}\xspace}                 
 \def\PUpsilon    {\ensuremath{\Upsilon}\xspace}
 \let\oldPi\Pi
 \def\PPi         {\ensuremath{\oldPi}\xspace}

 \def\PB      {\ensuremath{\mathrm{B}}\xspace}                 
 \def\PD      {\ensuremath{\mathrm{D}}\xspace}                 
 \def\PJ      {\ensuremath{\mathrm{J}}\xspace}                 
 \def\PK      {\ensuremath{\mathrm{K}}\xspace}                 
 \def\PP      {\ensuremath{\mathrm{P}}\xspace}                 
 \def\PS      {\ensuremath{\mathrm{S}}\xspace}                 

 \def\PW      {\ensuremath{\mathrm{W}}\xspace}                 
 \def\Pb      {\ensuremath{\mathrm{b}}\xspace}                 
 \def\Pc      {\ensuremath{\mathrm{c}}\xspace}                 
                  
 \def\Pe      {\ensuremath{\mathrm{e}}\xspace}                 
 \def\Pp      {\ensuremath{\mathrm{p}}\xspace}                 

 \def\Ps      {\ensuremath{\mathrm{s}}\xspace}

 \def\thebaroffset{0.0em}
}
{
 
 \def\Pgamma      {\ensuremath{\gamma}\xspace}

 \def\Pmu         {\ensuremath{\mu}\xspace}

 \def\Ppi         {\ensuremath{\pi}\xspace}

 \def\Pchi        {\ensuremath{\chi}\xspace}                 
 \def\Ppsi        {\ensuremath{\psi}\xspace}                 
                  
 \mathchardef\PDelta="7101
 \mathchardef\PXi="7104
 \mathchardef\PLambda="7103
 \mathchardef\PSigma="7106
 \mathchardef\POmega="710A
 \mathchardef\PUpsilon="7107
 \mathchardef\PPi="7105
 \def\PB      {\ensuremath{B}\xspace}                 
 \def\PD      {\ensuremath{D}\xspace}                 
 \def\PJ      {\ensuremath{J}\xspace}                 
 \def\PK      {\ensuremath{K}\xspace}                 
 \def\PP      {\ensuremath{P}\xspace}                 
 \def\PS      {\ensuremath{S}\xspace}                 
 \def\PW      {\ensuremath{W}\xspace}                 
 \def\Pb      {\ensuremath{b}\xspace}                 
 \def\Pc      {\ensuremath{c}\xspace}                 
                  
 \def\Pe      {\ensuremath{e}\xspace}                 
 \def\Pp      {\ensuremath{p}\xspace}                 

 \def\Ps      {\ensuremath{s}\xspace}

 \def\thebaroffset{0.18em}
}
\newcommand{\offsetoverline}[2][\thebaroffset]{\kern #1\overline{\kern -#1 #2}}%

\makeatletter
\ifcase \@ptsize \relax% 10pt
  \newcommand{\miniscule}{\@setfontsize\miniscule{4}{5}}% \tiny: 5/6
\or% 11pt
  \newcommand{\miniscule}{\@setfontsize\miniscule{5}{6}}% \tiny: 6/7
\or% 12pt
  \newcommand{\miniscule}{\@setfontsize\miniscule{5}{6}}% \tiny: 6/7
\fi
\makeatother

\DeclareRobustCommand{\optbar}[1]{\shortstack{{\miniscule (\rule[.5ex]{1.25em}{.18mm})}
  \\ [-.7ex] $#1$}}

\def\epem       {{\ensuremath{\Pe^+\Pe^-}}\xspace}

\def\mup        {{\ensuremath{\Pmu^+}}\xspace}
\def\mun        {{\ensuremath{\Pmu^-}}\xspace} % muon negative (\mum is taken)

\def\mumu       {{\ensuremath{\Pmu^+\Pmu^-}}\xspace}

\def\g      {{\ensuremath{\Pgamma}}\xspace}

\def\squark    {{\ensuremath{\Ps}}\xspace}

\def\cquark    {{\ensuremath{\Pc}}\xspace}

\def\bquark    {{\ensuremath{\Pb}}\xspace}

\def\pion   {{\ensuremath{\Ppi}}\xspace}

\def\pip    {{\ensuremath{\pion^+}}\xspace}
\def\pim    {{\ensuremath{\pion^-}}\xspace}

\def\kaon    {{\ensuremath{\PK}}\xspace}
\def\KorKbar {\kern \thebaroffset\optbar{\kern -\thebaroffset \PK}{}\xspace}

\def\Kp      {{\ensuremath{\kaon^+}}\xspace}

\def\D       {{\ensuremath{\PD}}\xspace}

\def\DorDbar {\kern \thebaroffset\optbar{\kern -\thebaroffset \PD}\xspace}

\def\Dp      {{\ensuremath{\D^+}}\xspace}
\def\Dm      {{\ensuremath{\D^-}}\xspace}

\def\DpDm    {\ensuremath{\Dp {\kern -0.16em \Dm}}\xspace}

\def\B       {{\ensuremath{\PB}}\xspace}

\def\BorBbar {\kern \thebaroffset\optbar{\kern -\thebaroffset \PB}\xspace}

\def\Bd      {{\ensuremath{\B^0}}\xspace}

\def\BdorBdbar {\kern \thebaroffset\optbar{\kern -\thebaroffset \Bd}\xspace}
\def\Bu      {{\ensuremath{\B^+}}\xspace}

\def\Bs      {{\ensuremath{\B^0_\squark}}\xspace}

\def\BsorBsbar {\kern \thebaroffset\optbar{\kern -\thebaroffset \Bs}\xspace}

\def\jpsi     {{\ensuremath{{\PJ\mskip -3mu/\mskip -2mu\Ppsi}}}\xspace}

\def\Y#1S{\ensuremath{\PUpsilon{(#1\PS)}}\xspace}
\def\OneS  {{\Y1S}\xspace}
\def\TwoS  {{\Y2S}\xspace}
\def\ThreeS{{\Y3S}\xspace}

\def\proton      {{\ensuremath{\Pp}}\xspace}

\def\LorLbar     {\kern \thebaroffset\optbar{\kern -\thebaroffset \PLambda}\xspace}

\newcommand{\decay}[2]{\ensuremath{#1\!\to #2}\xspace} 

\def\to                 {\ensuremath{\rightarrow}\xspace}

\def\AT#1     {\ensuremath{A_{\mathrm{T}}^{#1}}\xspace}           % 2

\def\C#1      {\ensuremath{\mathcal{C}_{#1}}\xspace}                       % 9
\def\Cp#1     {\ensuremath{\mathcal{C}_{#1}^{'}}\xspace}                    % 7
\def\Ceff#1   {\ensuremath{\mathcal{C}_{#1}^{\mathrm{(eff)}}}\xspace}        % 9  
\def\Cpeff#1  {\ensuremath{\mathcal{C}_{#1}^{'\mathrm{(eff)}}}\xspace}       % 7
\def\Ope#1    {\ensuremath{\mathcal{O}_{#1}}\xspace}                       % 2
\def\Opep#1   {\ensuremath{\mathcal{O}_{#1}^{'}}\xspace}                    % 7

\newcommand{\aunit}[1]{\ensuremath{\text{\,#1}}}       
\newcommand{\tev}{\aunit{Te\kern -0.1em V}\xspace}
\newcommand{\gev}{\aunit{Ge\kern -0.1em V}\xspace}
\newcommand{\mev}{\aunit{Me\kern -0.1em V}\xspace}
\newcommand{\kev}{\aunit{ke\kern -0.1em V}\xspace}
\newcommand{\ev}{\aunit{e\kern -0.1em V}\xspace}
 
\newcommand{\mevc}{\ensuremath{\aunit{Me\kern -0.1em V\!/}c}\xspace}
\newcommand{\gevc}{\ensuremath{\aunit{Ge\kern -0.1em V\!/}c}\xspace}
\newcommand{\kevcc}{\ensuremath{\aunit{ke\kern -0.1em V\!/}c^2}\xspace}
\newcommand{\mevcc}{\ensuremath{\aunit{Me\kern -0.1em V\!/}c^2}\xspace}
\newcommand{\gevcc}{\ensuremath{\aunit{Ge\kern -0.1em V\!/}c^2}\xspace}
\def\fb   {\ensuremath{\aunit{fb}}\xspace}
\def\invfb   {\ensuremath{\fb^{-1}}\xspace}

\newcommand{\chisq}{\ensuremath{\chi^2}\xspace}

\def\gsim{{~\raise.15em\hbox{$>$}\kern-.85em
          \lower.35em\hbox{$\sim$}~}\xspace}
\def\lsim{{~\raise.15em\hbox{$<$}\kern-.85em
          \lower.35em\hbox{$\sim$}~}\xspace}

\def\sPlot{\mbox{\em sPlot}\xspace}

\def\pt         {\ensuremath{p_{\mathrm{T}}}\xspace}

\def\ptot       {\ensuremath{p}\xspace}

\def\evtgen     {\mbox{\textsc{EvtGen}}\xspace}

\def\geant      {\mbox{\textsc{Geant4}}\xspace}

\def\photos     {\mbox{\textsc{Photos}}\xspace}

\def\tell1  {TELL1\xspace}
\def\ukl1   {UKL1\xspace}

\newcommand{\lhcborcid}[1]{\href{https://orcid.org/#1}{\hspace*{0.1em}\raisebox{-0.45ex}{\includegraphics[width=1em]{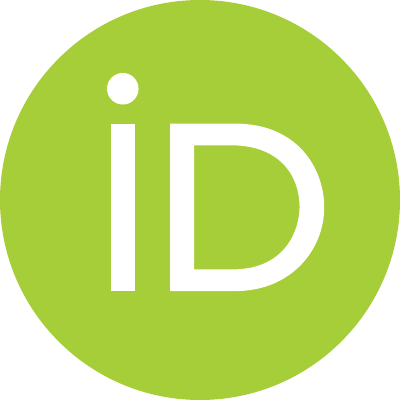}}}}

\hypersetup{
  colorlinks   = true, %Colours links instead of ugly boxes
  urlcolor     = blue, %Colour for external hyperlinks
  linkcolor    = blue, %Colour of internal links
  citecolor    = red   %Colour of citations
}

\ifEnableSectionTOCLinks
    \usepackage[explicit]{titlesec} % to change headings
    
    \let\oldcontentsline\contentsline
    \renewcommand

    \titleformat{\section}{\normalfont\Large\bf}{\hyperlink{tocsection.\thesection}{{\thesection} \parbox[t]{\dimexpr\textwidth-1pc}{#1}}}{1pc}{}

    \titleformat{\subsection}{\normalfont\bf}{\hyperlink{tocsubsection.\thesubsection}{{\thesubsection} \parbox[t]{\dimexpr\textwidth-1pc}{#1}}}{1pc}{}

    \titleformat{name=\section,numberless}[display]{}{}{0pt}{\normalfont\Huge\bfseries #1}
\fi

\usepackage{cite} % Allows for ranges in citations
\usepackage{mciteplus}
\usepackage{longtable} % only for template; not usually to be used in PAPERs

\def\pipi       {{\ensuremath{\pip\pim}}\xspace}
\begin{document}

%%%%%%%%%%%%%%%%%%%%%%%%%
%%%%% Title     %%%%%%%%%
%%%%%%%%%%%%%%%%%%%%%%%%%
\renewcommand{\thefootnote}{\fnsymbol{footnote}}
\setcounter{footnote}{1}

% %%%%%%% CHOOSE TITLE PAGE--------
%\onecolumn
%\input{title-LHCb-INT}
%\input{title-LHCb-ANA}
%\input{title-LHCb-CONF}
%\input{title-LHCb-FIGURE}
% ===============================================================================
% Purpose: LHCb-PAPER journal paper title page template
% Author: 
% Created on: 2010-09-25
% ===============================================================================

%%%%%%%%%%%%%%%%%%%%%%%%%
%%%%%  TITLE PAGE  %%%%%%
%%%%%%%%%%%%%%%%%%%%%%%%%
\begin{titlepage}
\pagenumbering{roman}

% Header ---------------------------------------------------
\vspace*{-1.5cm}
\centerline{\large EUROPEAN ORGANIZATION FOR NUCLEAR RESEARCH (CERN)}
\vspace*{1.5cm}
\noindent
\begin{tabular*}{\linewidth}{lc@{\extracolsep{\fill}}r@{\extracolsep{0pt}}}
\ifthenelse{\boolean{pdflatex}}% Logo format choice
{\vspace*{-1.5cm}\mbox{\!\!\!\includegraphics[width=.14\textwidth]{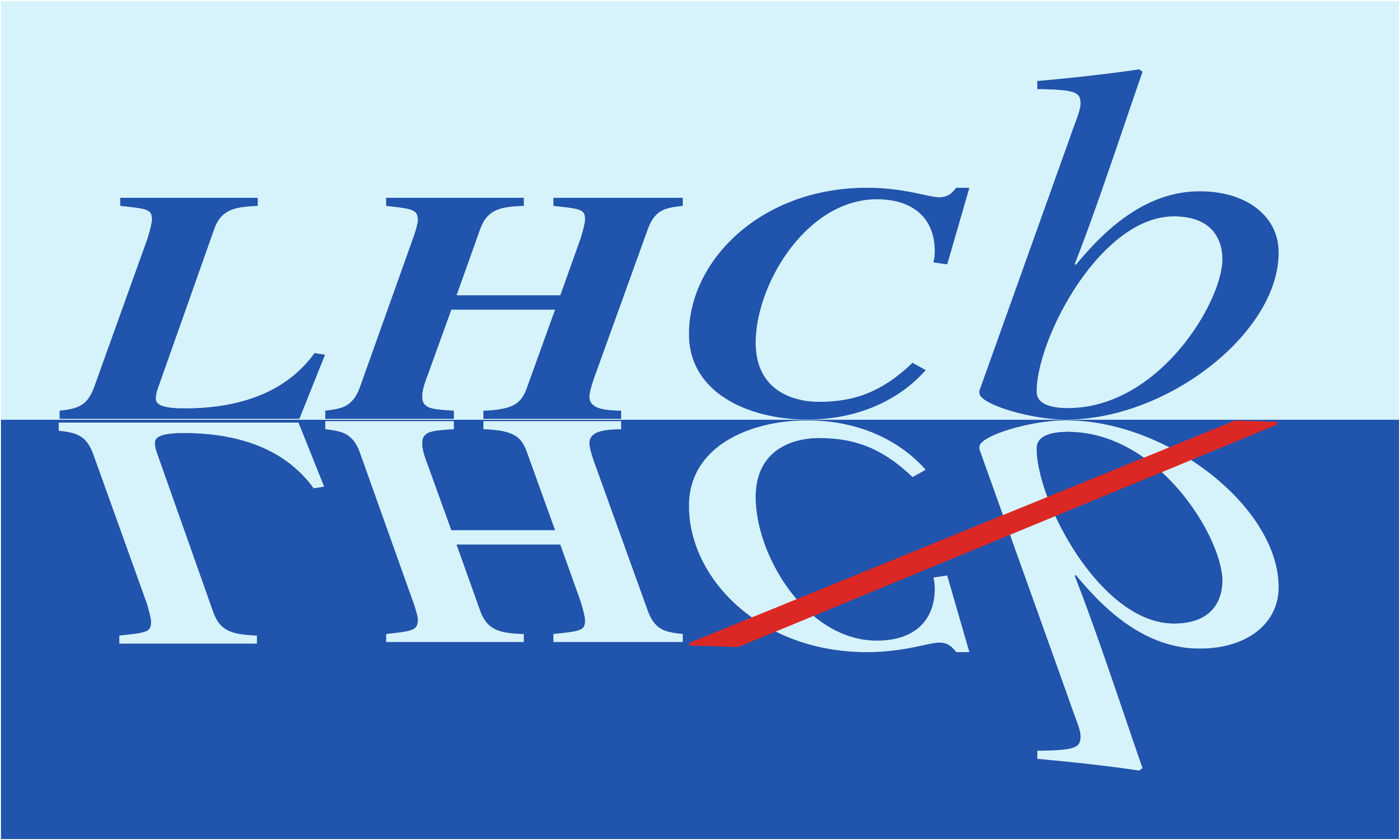}} & &}%
{\vspace*{-1.2cm}\mbox{\!\!\!\includegraphics[width=.12\textwidth]{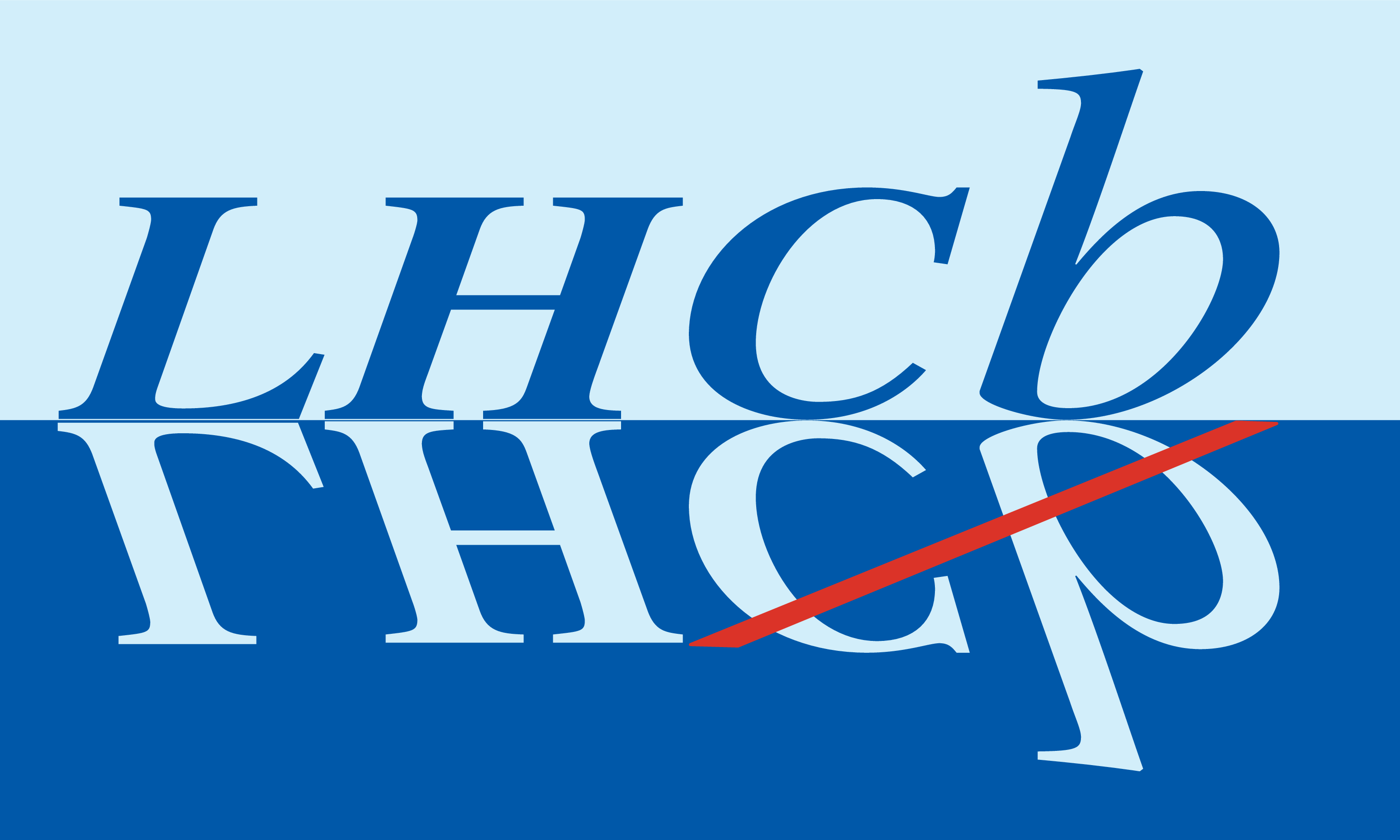}} & &}%
\\
 & & CERN-EP-2024-207 \\  % ID 
 & & LHCb-PAPER-2024-025 \\  % ID 
%% & & \today \\ % Date - Can also hardwire e.g.: 23 March 2010
& & August 8, 2024 \\ 
 %% & & v2.0 \\
% not in paper \hline
\end{tabular*}

\vspace*{1.5cm}

% Title --------------------------------------------------
{\normalfont\bfseries\boldmath\huge
\begin{center}
% DO NOT EDIT HERE. Instead edit macro in main.tex to keep metadata correct
  \papertitle 
\end{center}
}

\vspace*{1.0cm}

% Authors -------------------------------------------------
\begin{center}
%In the footnote, replace 'paper' by 'Letter' in case of submission to PRL or PLB 
% Edit macro in main.tex to keep metadata correct
\paperauthors\footnote{Authors are listed at the end of this paper.}
\end{center}

\vspace{\fill}

% Abstract -----------------------------------------------
\begin{abstract}
  \noindent
  The 
  decays of the $\Pchi_{\bquark1}(1\PP)$,
  $\Pchi_{\bquark2}(1\PP)$,
  $\Pchi_{\bquark1}(2\PP)$ and 
  $\Pchi_{\bquark2}(2\PP)$ mesons
  into the $\PUpsilon(1\PS)\mup\mun$ final state
  are observed with a high significance
  using proton-proton collision data collected with 
  the LHCb detector and corresponding to an integrated luminosity  of $9\invfb$.
  The newly observed decays
   together with the ${\PUpsilon(2\PS)}\rightarrow{\PUpsilon(1\PS)\pip\pim}$  and ${\PUpsilon(3\PS)}\rightarrow{\PUpsilon(2\PS)\pip\pim}$ decay modes
  are used for precision measurements of the mass and mass splittings 
  for the hidden-beauty states. 
\end{abstract}
%together with the $\decay{\PUpsilon(2\PS)}{\PUpsilon(1\PS)\pip\pim}$  and $\decay{\PUpsilon(3\PS)}{\PUpsilon(2\PS)\pip\pim}$ decay modes

\vspace*{1.0cm}

\begin{center}
  Published in \href{https://doi.org/10.1007/JHEP10(2024)122}{Journal of High Energy Physics 10 (2024) 122}
\end{center}

\vspace{\fill}

{\footnotesize 
% Edit macro in main.tex to keep metadata correct
\centerline{\copyright~\papercopyright. \href{\paperlicenceurl}{\paperlicence}.}}
\vspace*{2mm}

\end{titlepage}

%%%%%%%%%%%%%%%%%%%%%%%%%%%%%%%%
%%%%%  EOD OF TITLE PAGE  %%%%%%
%%%%%%%%%%%%%%%%%%%%%%%%%%%%%%%%

%  empty page follows the title page ----
\newpage
\setcounter{page}{2}
\mbox{~}
%\newpage
%

%\twocolumn
% %%%%%%%%%%%%% ---------

%% \listoftodos

\renewcommand{\thefootnote}{\arabic{footnote}}
\setcounter{footnote}{0}

%%%%%%%%%%%%%%%%%%%%%%%%%%%%%%%%
%%%%%  Table of Content   %%%%%%
%%%%%%%%%%%%%%%%%%%%%%%%%%%%%%%%
%%%% Uncomment if desired
%\tableofcontents

\cleardoublepage

%%%%%%%%%%%%%%%%%%%%%%%%%
%%%%% Main text %%%%%%%%%
%%%%%%%%%%%%%%%%%%%%%%%%%

\pagestyle{plain} % restore page numbers for the main text
\setcounter{page}{1}
\pagenumbering{arabic}

%% Uncomment during review phase. 
%% Comment before a final submission.
%% \linenumbers

%% This is the main body
%% It is useful to have a single file so comments are not missed in overleaf.
\section{Introduction}
\label{sec:Introduction}
Measurements of the properties of quarkonium states provide a test of QCD potential models~\cite{Eichten:1978tg,
Godfrey:2015dia,
Ferretti:2013vua,
Wang:2018rjg}. 
While the~charmonium system is precisely 
mapped out~\cite{Brambilla:2010cs,Gross:2022hyw}, 
the experimental knowledge of hidden beauty states is more limited. 
In Ref.~\cite{LHCb-PAPER-2017-036} it was shown that 
the~clean experimental
signature of $\Pchi_\cquark$ 
%% Dalitz 
decaying to the $\jpsi\mumu$~final state, 
referred to as muonic Dalitz decays hereafter, 
provides a new method to precisely measure properties of charmonia. 
In~this paper, the corresponding measurements in the~hidden beauty system are made. 
The~first observation of  the~muonic Dalitz decays of
the~$\Pchi_{\bquark1}(1\PP)$,
$\Pchi_{\bquark2}(1\PP)$, $\Pchi_{\bquark1}(2\PP)$,
and $\Pchi_{\bquark2}(2\PP)$~mesons 
to the~$\OneS$~state and measurement of the~masses of these states is reported.
These results have very different systematic uncertainties compared to those
obtained from the~measurement of the photon energy in $\TwoS$~or $\ThreeS$~transitions to 
the~$\Pchi_{\bquark1,2}(1\PP)$ or 
$\Pchi_{\bquark1,2}(2\PP)$~states~\cite{Lees:2014qea,
CLEO:2004jkt,
CrystalBall:1985pow,
ARGUS:1985qda,
Heintz:1992cv}.
%% %% BaBar:2014och
%% }.
Since these decay modes only involve muons, greater
precision and better control of systematics can be
achieved compared to studies with radiative modes 
involving
converted photons \cite{LHCb-PAPER-2014-040}.
In~addition, 
%% the~\mbox{$\decay{\ThreeS}{\TwoS \pipi}$} and 
%% \mbox{$\decay{\TwoS}{\OneS\pipi}$}~decay 
the~\mbox{$\decay{\ThreeS}{\left( \decay{\TwoS}{\mumu} \right) \pipi}$} and 
\mbox{$\decay{\TwoS}{\left(\decay{\OneS}{\mumu}\right)\pipi}$}~decay 
modes are used 
to make precise measurements of the~$\Upsilon$~meson masses 
and mass splittings.\footnote{In 
this paper, the~symbol $\PUpsilon$ 
denotes the~$\PUpsilon(1\PS)$,
$\PUpsilon(2\PS)$ and 
$\PUpsilon(3\PS)$~states together,
and the~symbol $\Pchi_\bquark$
stands for 
the~$\Pchi_{\bquark1}(1\PP)$,
$\Pchi_{\bquark2}(1\PP)$, 
$\Pchi_{\bquark1}(2\PP)$ and 
$\Pchi_{\bquark2}(2\PP)$~states.}
As~well as providing valuable input to QCD potential models, better knowledge 
of the~$\PUpsilon$~meson masses will improve the precision of electroweak 
measurements such as that of the~$\PW$~boson mass, where hidden beauty decays 
provide a standard method for detector calibration~\cite{LHCB-PAPER-2021-024}.

 This~analysis uses 
 LHCb data 
 collected
 in proton-proton\,($\proton\proton$) collisions 
 between 2011 and 2018.
 The~data from 2011 and 2012, collectively referred to as Run\,1, 
 were collected at centre\nobreakdash-of\nobreakdash-mass energies 
 of 7~and 8\tev and correspond to integrated luminosities 
 of 1\invfb and 2\invfb, respectively.
 The~rest of the~data, referred to as Run\,2,
 were collected between 
 2015 and 2018 at a~centre\nobreakdash-of\nobreakdash-mass energy 
 of 13\tev and correspond to an~integrated luminosity of~6\invfb.

%%%%%%%%%%%%%%%%%%%%%%%%%%%%%%%%%%%%%%%%%%%%%%%%%%%%%%%%%%%%%%%%%%%%%%%%%%%%%%%%%%%%%%%%%%%%%%%%%
\section{Detector and simulation}
\label{sec:Detector}
The \lhcb detector~\cite{LHCb-DP-2008-001,LHCb-DP-2014-002} is a single-arm forward spectrometer
that covers the~\mbox{pseudorapidity} range $2<\eta <5$, designed for the study of particles
containing \bquark or \cquark quarks. It~includes a high-precision tracking system consisting 
of a silicon-strip vertex detector  %% \,(VELO)
surrounding the $\proton\proton$~interaction region, 
a large-area silicon-strip detector %% \,(TT) 
located upstream of a dipole magnet with 
a bending power of approximately $4{\mathrm{\,T\,m}}$, and three stations of 
silicon-strip detectors and straw drift tubes  placed downstream of the~magnet. 
The~tracking system provides a measurement of the momentum, \ptot,  
of charged particles with a relative uncertainty that varies from~0.5\% at low momentum  
to~1.0\% at $200\gevc$. 
Large samples of  $\decay{\jpsi}{\mumu}$~and $\decay{\Bu}{\jpsi\Kp}$~decays,
collected concurrently with the~dataset used in this analysis, 
are used to calibrate the momentum scale of the spectrometer~\cite{LHCb-DP-2023-003}.
The relative uncertainty achieved on the momentum scale is $3 \times 10^{-4}$. 
Charged hadrons are distinguished using information from two ring\nobreakdash-imaging 
Cherenkov\,(RICH) detectors. In addition, photons, electrons, and hadrons are 
identified by a calorimeter system consisting of scintillating-pad and preshower 
detectors, an electromagnetic and a hadronic calorimeter.
%% The calorimeter response is calibrated using samples of $\decay{\piz}{\g\g}$~decays~\cite{LHCb-DP-2020-001}. 
Muons are identified by a system composed of alternating 
layers of iron and multiwire proportional chambers~\cite{LHCb-TDR-004,LHCb-DP-2012-002}.

The online event selection is performed by a~trigger, 
which consists of a~hardware system followed
by a~two\nobreakdash-level software stage~\cite{LHCb-DP-2012-004,LHCb-DP-2019-001}.
At~the~hardware trigger stage, events are required to have 
either a~muon track with high transverse momentum or 
a~dimuon candidate with a~high value for the~product of 
the~\pt of the~muons. For~the~Run\,2 dataset, 
the~alignment  and calibration of the~detector is performed 
in near real-time such that the results are used in the software trigger~\cite{Borghi:2017hfp}. 
The same alignment and calibration information 
is propagated to the offline reconstruction, ensuring consistent information between the trigger 
and offline software. The first level of the software trigger performs a partial event reconstruction 
and requires events to have a~pair of well-identified oppositely charged muons with 
a~reconstructed mass larger 
than $2.7 \gevcc$.  %%  without biasing the lifetime distribution. 
The second level performs a 
full event reconstruction. Events are retained for further 
processing if they contain a~high-mass dimuon %$\decay{\PUpsilon}{\mumu}$
candidate.

This analysis makes limited use of simulation. The main input needed from simulation is 
the~resolution model 
for the~mass distribution of the signal decays. 
This~is obtained using the~{\sc{RapidSim}} fast-simulation package~\cite{Cowan:2016tnm}. 
This has been tuned using the LHCb full simulation,
implemented with the \geant toolkit~\cite{Agostinelli:2002hh,Allison:2006ve} 
as described in Ref.~\cite{LHCb-PROC-2011-006}. 
In~both simulation frameworks, 
decays of hadronic particles are described by \evtgen~\cite{Lange:2001uf},
in which final-state radiation is generated using \photos~\cite{Golonka:2005pn,davidson2015photos}. 
The~\mbox{$\decay{\PUpsilon(2\PS)}{\PUpsilon(1\PS)\pipi}$}~decay 
%% is modelled using 
%% the {\tt VVPPIPI} model,  
follows the~model described in 
 Refs.~\cite{Brown:1975dz,Voloshin:2006ce,Voloshin:2007dx}, 
the~\mbox{$\decay{\PUpsilon(3\PS)}{\PUpsilon(2\PS)\pipi}$} decay
uses  
a~phase\nobreakdash-space distribution, 
and the~Dalitz decays
are based on the~approach described in Ref.~\cite{Luchinsky:2017pby}. 
The $\pt$ spectrum for hidden beauty mesons is taken from the LHCb studies of $\PUpsilon$ production presented 
in Refs.~\cite{LHCb-PAPER-2011-036, 
LHcb-PAPER-2013-016, 
LHCb-PAPER-2013-066, 
LHCb-PAPER-2015-045, 
LHCb-PAPER-2018-002}.

\section{$\decay{\PUpsilon}{\mumu}$ selection and mass fits}
\label{sec:mumu}
The selection starts from a~pair of oppositely charged muons,
%% identified by the trigger system 
%% as consistent with being produced in a~\mbox{$\decay{\PUpsilon}{\mumu}$}~decay. 
selected by the~trigger as an~\mbox{$\decay{\PUpsilon}{\mumu}$}~candidate.
Combinatorial background is efficiently 
rejected by a~requirement on the~output of a~boosted decision tree classifier~\cite{Breiman,AdaBoost},
referred to as {\sc{BDT}}~hereafter, 
implemented in the~{\sc{TMVA}}~toolkit~\mbox{\cite{Hocker:2007ht,TMVA4}}. 
The~{\sc{BDT}}~classifier is trained on a~small fraction of the~data 
using the~\sPlot~technique~\cite{Pivk:2004ty}, with the dimuon mass as the discriminating variable,  
to distinguish signal and background.
As~inputs, 
the~classifier uses seventeen variables related
to the decay kinematics, 
track quality~\cite{LHCb-DP-2013-002,LHCb-DP-2013-003,LHCb-DP-2017-001,DeCian:2255039},
vertex quality and particle identification information~\cite{LHCb-PROC-2011-008,
LHCb-DP-2012-002}.
A~loose requirement 
on the~{\sc{BDT}} output reduces the~combinatorial background by 
a~factor of four, whilst only removing $1\%$ of the~signal. 

\begin{figure}[t]
  \setlength{\unitlength}{1mm}
  \centering
  \definecolor{gr}{rgb}{0.35, 0.83, 0.33}
  \definecolor{br}{rgb}{0.43, 0.98, 0.98}
  \definecolor{vi}{rgb}{0.39, 0.37, 0.96}
  \definecolor{db}{rgb}{0.1, 0.08, 0.41}
  \begin{picture}(150,120)
   \put(  0,  0){ 
      \includegraphics*[width=150mm,%height=60mm,%
      %%]{Y_MVA_All.pdf}
      ]{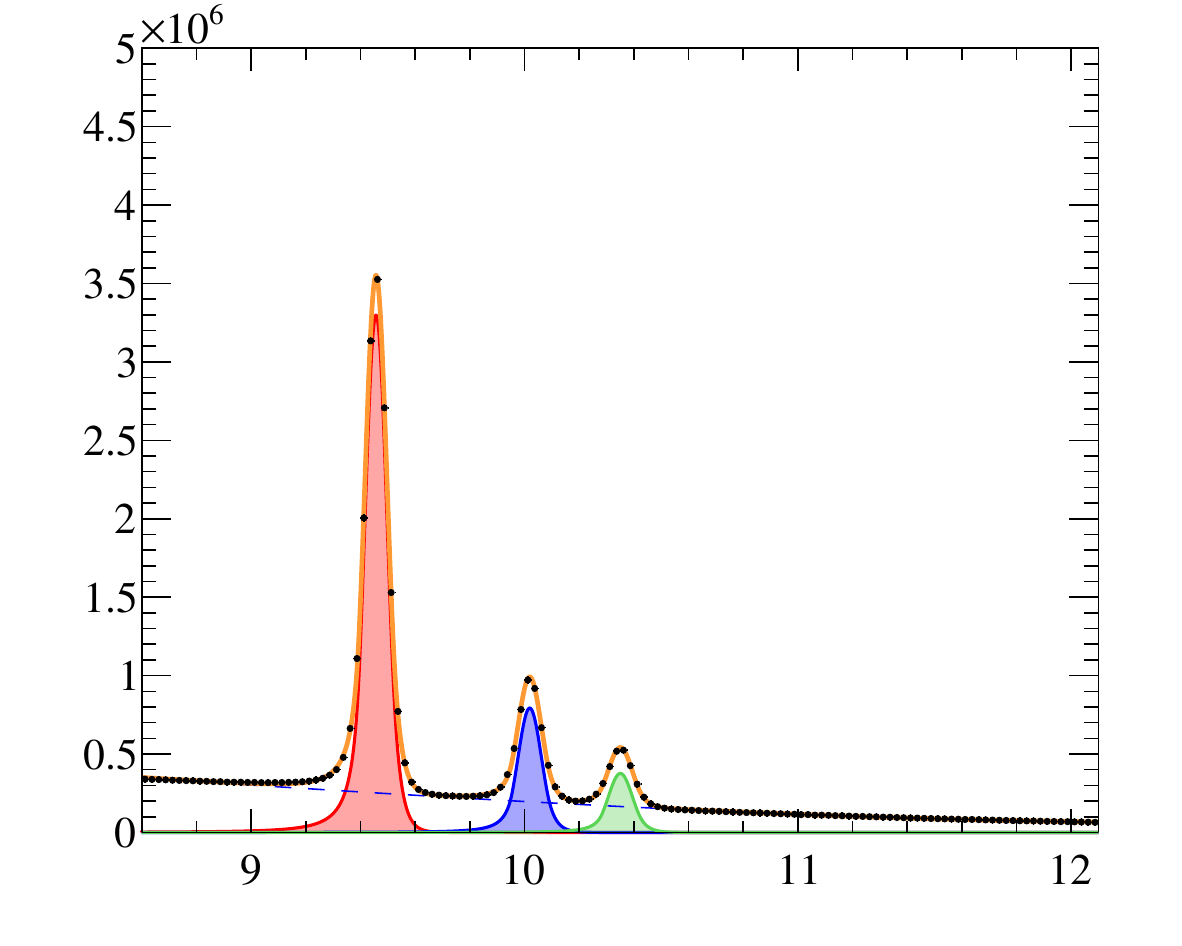}
    } 
    \put(25,100){\Large$\begin{array}{l}\lhcb\\9\invfb\end{array}$}
     \put(92,90){$\begin{array}{cl}
	 %% DATA 
	 \!\!\!\bigplus\!\!\!\!\!\!\bullet\mkern-5mu&\mathrm{Data}  
	 \\ 
	 \begin{tikzpicture}[x=1mm,y=1mm]\filldraw[fill=red!35!white,draw=red,thick]  (0,0) rectangle (10,4);\end{tikzpicture} 
      & \decay{\PUpsilon(1\PS)}{\mumu} 
	 \\
	 \begin{tikzpicture}[x=1mm,y=1mm]\filldraw[fill=blue!35!white,draw=blue,thick]  (0,0) rectangle (10,4);\end{tikzpicture} 
      & \decay{\PUpsilon(2\PS)}{\mumu} 
      \\
	 \begin{tikzpicture}[x=1mm,y=1mm]\filldraw[fill=gr!35!white,draw=gr,thick]  (0,0) rectangle (10,4);\end{tikzpicture} 
      & \decay{\PUpsilon(3\PS)}{\mumu} 
	 \\
      {\color[RGB]{0,0,255}{\hdashrule[0.5ex][x]{10mm}{1.3pt}{2.3mm 0.8mm}}} & \mathrm{Background}
	 \\
      {\color[RGB]{255,153,51} {\rule[0.5ex]{10mm}{2.0pt}}} & \mathrm{Total}
	 \end{array}$}
    \put( 75,1){\Large$m_{\mumu}$}     \put(120,1){\Large$\left[\!\gevcc\right]$}  
    \put(3,56){\Large\begin{sideways}Candidates/($25\mevcc$)\end{sideways}}
  \end{picture}
  \caption { \small
  Dimuon mass spectrum
  after the {\sc{BDT}}~requirement. 
  The~result of the~fit described in the~text is overlaid.}
  \label{fig:mumu}
\end{figure}
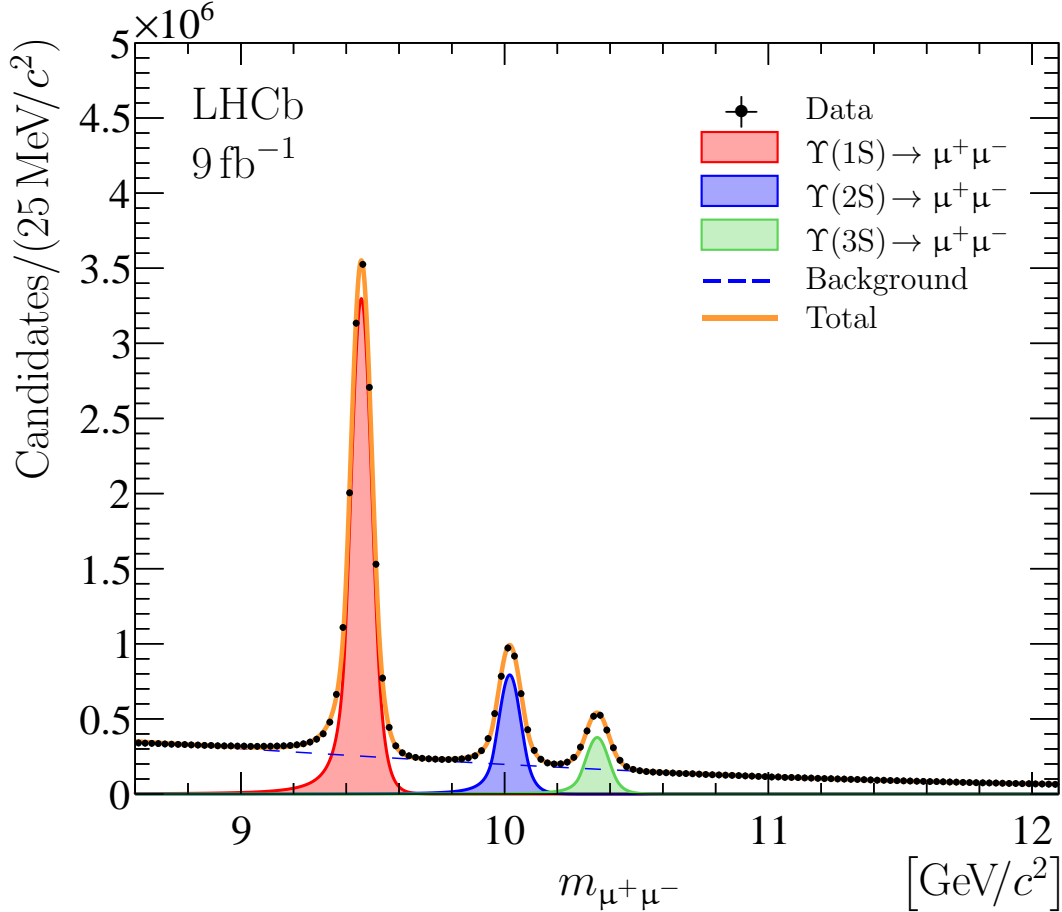

The~dimuon mass distribution obtained after the~{\sc{BDT}} requirement 
is shown in Fig.~\ref{fig:mumu}.
A fit is made to this distribution to extract the mass parameters. 
In this fit 
the~\mbox{$\decay{\PUpsilon(1\PS)}{\mumu}$}, 
\mbox{$\decay{\PUpsilon(2\PS)}{\mumu}$} 
and 
\mbox{$\decay{\PUpsilon(3\PS)}{\mumu}$}~signals
are  described by 
modified Gaussian functions with power-law tails on both 
sides of the distribution~\cite{Skwarnicki:1986xj,LHCb-PAPER-2011-013}.
The~tail parameters of the~modified Gaussian functions are shared
between the~three signal peaks,
and allowed to vary in the fit,  
%% and the~widths of the~Gaussian 
while their widths 
are constrained to scale linearly 
with the~dimuon mass~\mbox{\cite{LHCb-PAPER-2011-036,
LHcb-PAPER-2013-016,
LHCb-PAPER-2013-066,
LHCb-PAPER-2015-045}}.
A~third\nobreakdash-order 
polynomial is used to describe the~background.  
Due~to the~very large data sample a~binned extended maximum-likelihood fit is used. 
The~bin width, of 1\mevcc, is chosen to be sufficiently narrow so that it does not affect 
the~determination of the~$\PUpsilon$~masses.
For~the~considered fit range  
the~efficiency is found to be 
a~smooth and slowly varying
function of the~dimuon mass, 
not affecting the~mass measurements.
The~measured masses and mass differences are corrected for the~effect of QED radiative 
processes using  %  the bias obtained from 
simulation. 
The~correction for the~$\PUpsilon(1\PS)$~mass is large, 3.27\mevcc. 
It largely cancels in the~mass~splittings, resulting 
in residual corrections of $0.29$ and $0.14\mevcc$ 
for the~\mbox{$m_{\PUpsilon(2\PS)}-m_{\PUpsilon(1\PS)}$}
and \mbox{$m_{\PUpsilon(3\PS)}-m_{\PUpsilon(2\PS)}$}~differences, 
respectively. 

\begin{table}[b]
\centering
\caption{\small
Parameters of interest,
yields $N_{\PUpsilon}$, 
masses and mass differences, 
from the~fit to 
the~dimuon mass spectra. The uncertainties are statistical only.} 
%% \begin{scriptsize}
\label{tab:Y_fits}
	\vspace{2mm}
	\begin{tabular*}{0.60\textwidth}
	{@{\hspace{5mm}}l@{\extracolsep{\fill}}lc@{\hspace{5mm}}}
 \multicolumn{2}{l}{Parameter} 
 &   Value 
   \\[1.5mm]
  \hline 
  \\[-1.5mm]
  $N_{\PUpsilon(1\PS)}$
  & $\left[10^3\right]$
  & $14\,609.3  \pm 6.7 $  
  \\
  $N_{\PUpsilon(2\PS)}$
  & $\left[10^3\right]$
  & $\phantom{0}3\,729.1 \pm 3.2$  
  \\
  $N_{\PUpsilon(3\PS)}$
  & $\left[10^3\right]$
  & $\phantom{0}1\,827.9 \pm 2.3$  
  \\
  $m_{\PUpsilon(1\PS)}$
  &  $\left[\!\mevcc\right]$ 
  &   $\phantom{0}9\,460.37 \pm  0.01$  
  %% &   $\phantom{0}9\,457.097 \pm  0.014$ 
  %% m_{\OneS} & = & 9\,460.37 \pm \, 0.01 \stat \pm 2.85 \syst  \, \mevcc\,, \label{eq:res_mumu_1s}
  \\
  $m_{\PUpsilon(2\PS)}-m_{\PUpsilon(1\PS)}$
  &  $\left[\!\mevcc\right]$ 
  &  $\phantom{00\,}562.71 \pm 0.04$
  %% &  $\phantom{00\,}562.415 \pm 0.037$
  %% m_{\TwoS}  - m_{\OneS}  & = & \phantom{0\,}562.71 \pm \, 0.04 \stat \pm 0.30 \syst \, \mevcc \,,
  %%  
  \\
   $m_{\PUpsilon(3\PS)}-m_{\PUpsilon(2\PS)}$
  &  $\left[\!\mevcc\right]$ 
  &  $\phantom{00\,}331.77 \pm 0.07$
  %% &  $\phantom{00\,}331.633 \pm 0.067$ 
  %% m_{\ThreeS}  - m_{\TwoS}  & = & \phantom{0\,}331.77 \pm \, 0.067 \stat \pm 0.10 \syst  \, \mevcc
  %%  
\end{tabular*}
%% \end{scriptsize}
\end{table}

The large yields for the~\mbox{$\decay{\PUpsilon}{\mumu}$}~modes
and the~low\nobreakdash-background level, lead to a small statistical uncertainty
on the mass parameters, given in Table~\ref{tab:Y_fits}. 
The~dominant systematic uncertainty on the~$\OneS$ mass, 
$2.8 \mevcc$, is due to the~knowledge of
the~momentum scale~\cite{LHCb-DP-2023-003}.
%% \,\mbox{($\pm 0.3 \times 10^{-3}$)}~\cite{LHCb-DP-2023-003}.
In~the~mass differences, this~uncertainty largely cancels, 
giving an uncertainty of $0.17 \mevcc$  for the $m_{\PUpsilon(2\PS)}-m_{\PUpsilon(1\PS)}$ 
and $0.10 \mevcc$ for the $m_{\PUpsilon(3\PS)}-m_{\PUpsilon(2\PS)}$~splitting.
An~additional uncertainty arises from the~size of
the~energy-loss correction applied in the~track fit~\cite{LHcb-PAPER-2013-011}. 
Varying the~detector material within its 10\% uncertainty and rerunning 
the~track fit results in a $20\kevcc$~uncertainty for two\nobreakdash-body decay modes.
Further uncertainties arise from the knowledge of the radiative 
corrections and the assumed fit model. 
The systematic uncertainty from the~modelling of the~former
is estimated by running \photos with 
different settings \cite{LHCb-DP-2023-003,LHCb-PAPER-2011-035} and largely 
cancels in the mass differences. The~uncertainty on the~latter is evaluated using alternative fit models for
the~background, which include convex decreasing polynomials of order two, three, or four; 
generic polynomials of the second and fourth order; and a~product of an~exponential function 
and a~first-order polynomial function. As~an~alternative signal shape, 
a modified Gaussian function with tail parameters constrained 
to values obtained in~previous LHCb analyses~\cite{LHCb-PAPER-2011-036,
LHcb-PAPER-2013-016,
LHCb-PAPER-2013-066,
LHCb-PAPER-2014-015,
LHCb-PAPER-2015-045,
LHCb-PAPER-2015-046,
LHCb-PAPER-2017-028,
LHCb-PAPER-2018-002} is considered.
The total systematic uncertainties for the mass and mass~difference measurements
are summarised in Table~\ref{tab:mumu_syst}.

\begin{table}[tb]
\centering
\caption{\small Systematic uncertainties 
on the measurement of
the~$\Upsilon(1\PS)$~mass 
and mass differences from 
the~analysis 
of the~dimuon mass spectrum.}
\label{tab:mumu_syst}
	\vspace{2mm}
	\begin{tabular*}{0.85\textwidth}
	{@{\hspace{5mm}}l@{\extracolsep{\fill}}ccc@{\hspace{5mm}}}
\multirow{2}{*}{Source of systematic} 
  & \multicolumn{3}{c} {Uncertainty $\left[\!\mevcc\right]$} \\
  & $m_{\OneS}$ 
  & $m_{\PUpsilon(2\PS)}-m_{\PUpsilon(1\PS)}$  
  & $m_{\PUpsilon(3\PS)}-m_{\PUpsilon(2\PS)}$ 
   \\[1.5mm]
  \hline 
  \\[-1.5mm]
Momentum scale              &   2.8\phantom{0}   &  0.17    & 0.10  \\
Energy loss correction      &   0.02             &   ---    &  ---  \\
Radiative corrections       &   0.13             &  0.03    & 0.03  \\
Fit model                   &   0.35             &  0.10    & 0.02   
   \\[1.5mm]
  \hline 
  \\[-1.5mm]
Sum in quadrature  & 2.85 & 0.20 &  0.10  \\
% 0.11, 0.07
\end{tabular*}
\end{table}

\section{$\decay{\PUpsilon(2\PS)}{\PUpsilon(1\PS)\pipi}$
and $\decay{\PUpsilon(3\PS)}{\PUpsilon(2\PS)\pipi}$ selection and mass fits} 
\label{sec:Ups_pipi}

The~\mbox{$\decay{\PUpsilon}{\mumu}$}~candidates 
with mass in 
the~regions 
\mbox{$9.343<m_{\mumu}<9.556\gevcc$} and 
\mbox{$9.899<m_{\mumu}<10.124\gevcc$} 
are considered as 
\mbox{$\decay{\PUpsilon(1\PS)}{\mumu}$} and 
\mbox{$\decay{\PUpsilon(2\PS)}{\mumu}$}~candidates, respectively. 
Each mass region contains 
95\% of the corresponding decays, according to 
the~fit described in Sec.~\ref{sec:mumu}.
To form \mbox{$\decay{\PUpsilon(2\PS)}{\PUpsilon(1\PS)\pipi}$} 
and \mbox{$\decay{\PUpsilon(3\PS)}{\PUpsilon(2\PS)\pipi}$}~candidates, 
selected \mbox{$\decay{\PUpsilon(1\PS)}{\mumu}$}
and \mbox{$\decay{\PUpsilon(2\PS)}{\mumu}$} candidates 
are combined with pairs of oppositely charged particles 
identified as pions by the RICH system~\cite{LHCb-PROC-2011-008,LHCb-DP-2012-003}.
Combinatorial background is reduced by requiring 
the~scalar sum of the~\pt of the~pions to be larger than $400 \mevc$.
To~further suppress background,
a~parameter $x_{\pipi}$
is introduced as 
 \begin{equation}
 \label{eq:x23}
        x_{\pipi} \equiv \dfrac{ m_{\pipi} - 2 m_{\pion}}
        { m_{\PUpsilon\pipi} - m_{\PUpsilon} - 2m_{\pion}} \,,
    \end{equation}
where     
$m_{\pipi}$, 
$m_{\PUpsilon}$ 
and $m_{\PUpsilon\pipi}$  
stand for  the masses of 
the~$\pipi$, 
$\mumu$
and $\PUpsilon\pipi$~systems,
and 
$m_{\pion}$~is the known mass of the charged pion~\cite{PDG2024}.
Candidates are 
required to satisfy a loose requirement, $x_{\pipi} > 0.2$. 
A~kinematic fit~\cite{Hulsbergen:2005pu} is made to the selected candidates, 
constraining the~dimuon mass to 
the~known $\OneS$~or $\TwoS$~mass~\cite{PDG2024}, 
indicated in Table~\ref{tab:umasspdg}, 
as appropriate, 
and requiring the candidate to be consistent with 
coming from a~primary \proton\proton collision vertex. 
The~$\chisq$ per degree of freedom of this fit, 
$\chisq_{\mathrm{fit}}/{\mathrm{ndf}}$,  is required to be less than five.

\begin{figure}[t]
  \setlength{\unitlength}{1mm}
  \centering
  \definecolor{gr}{rgb}{0.35, 0.83, 0.33}
  \definecolor{br}{rgb}{0.43, 0.98, 0.98}
  \definecolor{vi}{rgb}{0.39, 0.37, 0.96}
  \definecolor{db}{rgb}{0.1, 0.08, 0.41}
  \begin{picture}(150,120)    
  \put(   0,   0){ 
      \includegraphics*[width=150mm,%height=60mm,%
      %% ]{Y2S_pipi.pdf}
      ]{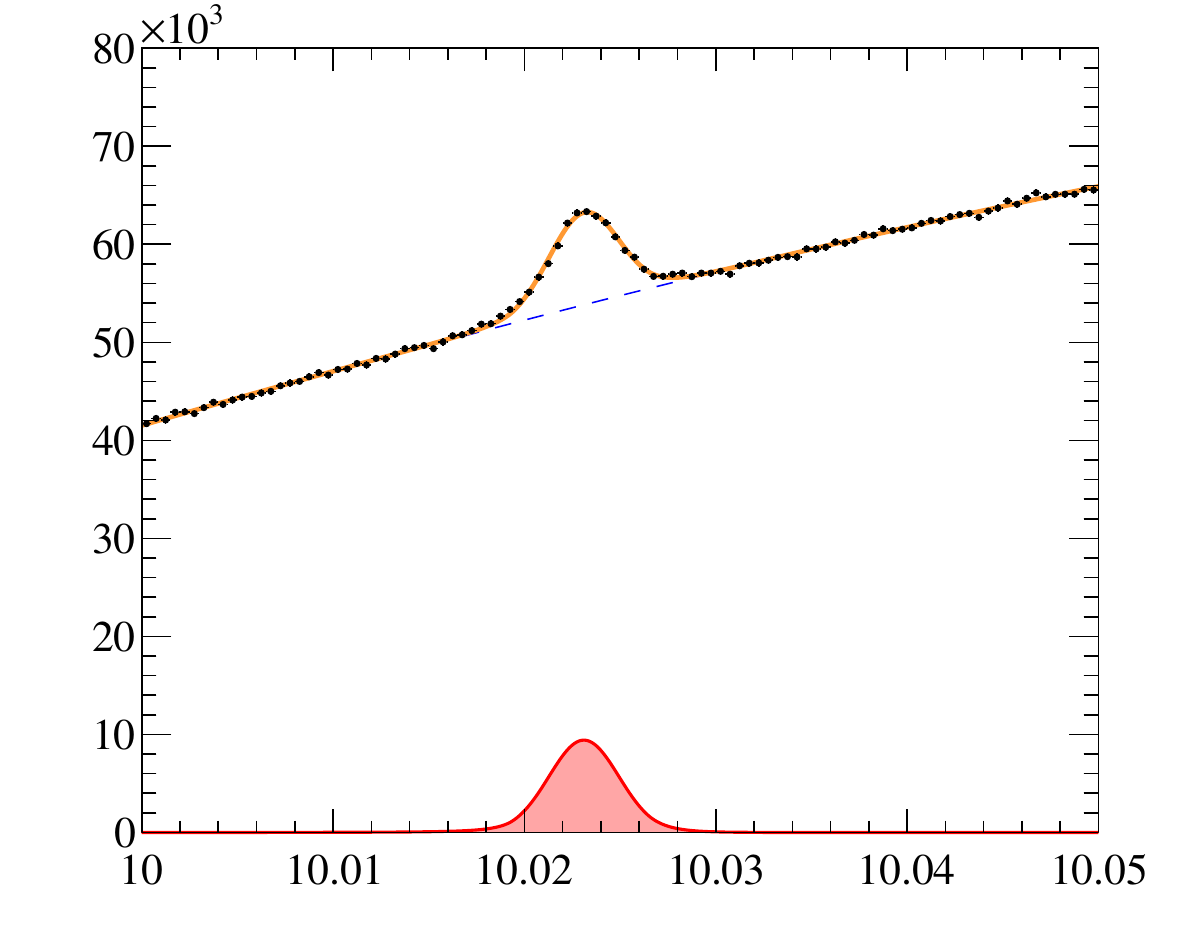}
    }
    \put(115,100){\Large$\begin{array}{l}\lhcb\\9\invfb\end{array}$}
     \put(25,98){$\begin{array}{cl}
	 %% DATA 
	 \!\!\!\bigplus\!\!\!\!\!\!\bullet\mkern-5mu&\mathrm{Data}  
	 \\ 
	 \begin{tikzpicture}[x=1mm,y=1mm]\filldraw[fill=red!35!white,draw=red,thick]  (0,0) rectangle (10,4);\end{tikzpicture} 
      & \decay{\PUpsilon(2\PS)}{\PUpsilon(1\PS)\pipi} 
	 \\
      {\color[RGB]{0,0,255}{\hdashrule[0.5ex][x]{10mm}{1.3pt}{2.3mm 0.8mm}}} & \mathrm{Background}
	 \\
      {\color[RGB]{255,153,51} {\rule[0.5ex]{10mm}{2.0pt}}} & \mathrm{Total}
	 \end{array}$}
    \put(75, 0){\Large$m_{\PUpsilon(1\PS)\pip\pim}$}  \put(120,0){\Large$\left[\!\gevcc\right]$}
    \put( 3,55){\Large\begin{sideways}Candidates/$(0.5\mevcc)$\end{sideways}}
  \end{picture}
  \caption { \small
  Mass spectrum of $\PUpsilon(1\PS)\pip\pim$~candidates 
  with a constraint on the  $\PUpsilon(1\PS)$ mass applied. The~result of the~fit described in the~text is overlaid.
  }
  \label{fig:UpsTwo}
\end{figure}

\begin{figure}[t]
  \setlength{\unitlength}{1mm}
  \centering
  \definecolor{gr}{rgb}{0.35, 0.83, 0.33}
  \definecolor{br}{rgb}{0.43, 0.98, 0.98}
  \definecolor{vi}{rgb}{0.39, 0.37, 0.96}
  \definecolor{db}{rgb}{0.1, 0.08, 0.41}
  \begin{picture}(150,120)
   \put(   0,   0){ 
      \includegraphics*[width=150mm,%height=60mm,%
     %%]{Y3S_pipi.pdf}
      ]{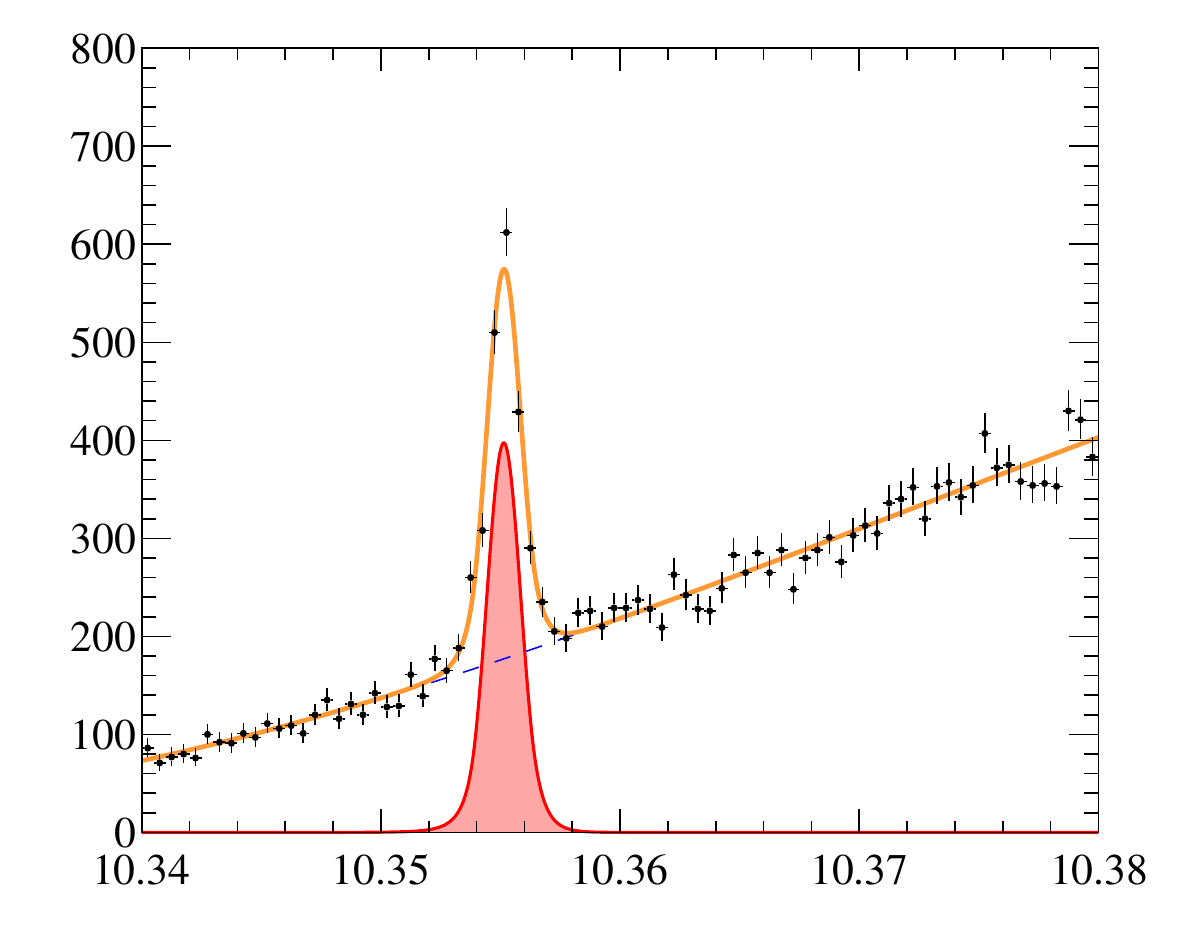}
    }
    \put(115,100){\Large$\begin{array}{l}\lhcb\\9\invfb\end{array}$}
     \put(25,100){$\begin{array}{cl}
	 %% DATA 
	 \!\!\!\bigplus\!\!\!\!\!\!\bullet\mkern-5mu&\mathrm{Data}  
	 \\ 
	 \begin{tikzpicture}[x=1mm,y=1mm]\filldraw[fill=red!35!white,draw=red,thick]  (0,0) rectangle (10,4);\end{tikzpicture} 
      & \decay{\PUpsilon(3\PS)}{\PUpsilon(2\PS)\pipi} 
	 \\
      {\color[RGB]{0,0,255}{\hdashrule[0.5ex][x]{10mm}{1.3pt}{2.3mm 0.8mm}}} & \mathrm{Background}
	 \\
      {\color[RGB]{255,153,51} {\rule[0.5ex]{10mm}{2.0pt}}} & \mathrm{Total}
	 \end{array}$}
    \put(75, 0){\Large$m_{\PUpsilon(2\PS)\pip\pim}$}  \put(120,0){\Large$\left[\!\gevcc\right]$}
    \put( 3,55){\Large\begin{sideways}Candidates/$(0.5\mevcc)$\end{sideways}}
  \end{picture}
  \caption { \small
  Mass spectrum of $\PUpsilon(2\PS)\pip\pim$~candidates
  with a constraint on the $\PUpsilon(2\PS)$ mass applied. 
  The~result of the~fit described in the~text is overlaid.
    }
  \label{fig:UpsThree}
\end{figure}

The $\PUpsilon(1\PS)\pipi$ and $\PUpsilon(2\PS)\pipi$
mass spectra are shown in Figs.~\ref{fig:UpsTwo} and~\ref{fig:UpsThree}, respectively.
A~fit to the~$\PUpsilon(1\PS)\pipi$~mass spectrum
is performed using a~function consisting 
of signal and combinatorial background components.
The~signal component is described using the~modified Gaussian function described 
in Sec.~\ref{sec:mumu}, 
%% above, 
with all parameters fixed from the~simulation
apart from the~peak location
and a~scale factor $s_f$ for the~width,
which allows for differences between data and 
simulation~\cite{LHCb-PAPER-2020-008,
LHCb-PAPER-2020-009,
LHCb-PAPER-2020-035,
LHCb-PAPER-2021-031,
LHCb-PAPER-2021-032}.
The~background component is modelled by 
an~increasing third\nobreakdash-order polynomial function.
%% To account for~potential differences between the data and the simulation,
%% the resolution parameter is scaled with a factor, $s_f$, that is allowed to vary. 
%% 
An~extended binned maximum-likelihood fit is used 
with the~bin width  of $100\kevcc$ chosen to be 
sufficiently narrow so that it does not affect 
the~$\TwoS$~mass  determination.
The~fit result is overlaid in Fig.~\ref{fig:UpsTwo} and the~results for 
the parameters of interest are listed in Table~\ref{tab:Ypipi_fits}.
The~mass results in Table~\ref{tab:Ypipi_fits} 
include 
%% the effect of radiative corrections.
a correction for QED radiation. 
%% The~value of $s_f$ is found to be consistent with unity   
%% \begin{equation}  \label{eq:sf_twos}
%%     s_f = 1.015 \pm  0.012 \,. 
%% \end{equation}
The~value of $s_f= 1.015 \pm  0.012$  is found to be consistent with unity.   
%%  \begin{equation}  \label{eq:sf_twos}
%%     s_f = 1.015 \pm  0.012 \,. 
%% \end{equation}

A similar signal model, obtained from simulation,
is used to fit the~$\PUpsilon(2\PS)\pipi$~mass spectrum. 
In~this case 
the~background component
is described by an~increasing convex third\nobreakdash-degree polynomial function.
For this fit, the~resolution 
scale factor is  Gaussian\nobreakdash-constrained to the~value obtained 
for the~$\decay{\PUpsilon(2\PS)}{\PUpsilon(1\PS)\pip\pim}$~signal
above. 
%% in Eq.~\eqref{eq:sf_twos}. 
The~result of an~extended unbinned maximum-likelihood fit
%%
%% The~fit result
is overlaid in Fig.~\ref{fig:UpsThree}
and the~results for the~parameters of interest are listed in Table~\ref{tab:Ypipi_fits}.
The~combined resolution scale factor determined 
from the~fits 
to~the~$\decay{\PUpsilon(2\PS)}{\PUpsilon(1\PS)\pip\pim}$
and the~$\decay{\PUpsilon(3\PS)}{\PUpsilon(2\PS)\pip\pim}$
signals is 
\begin{equation} \label{eq:scale_factor}
    s_f = 1.012 \pm  0.012 \,. 
\end{equation}
%% 
%% are corrected for 
%% effect of the~radiative corrections.
%%
%%
\begin{table}[tb]
\centering
\caption{\small
Values for the~parameters of interest,
yields $N$ and masses,  
from the~fits
to the~$\PUpsilon(1\PS)\pipi$
and $\PUpsilon(2\PS)\pipi$~mass spectra.
%% from Figs.~\ref{fig:mva:upstwo} and~\ref{fig:mva:upsthree}.
Uncertainties  are statistical only.
    } 
	\label{tab:Ypipi_fits}
	\vspace{2mm}
	\begin{tabular*}{0.60\textwidth}
	{@{\hspace{5mm}}l@{\extracolsep{\fill}}lc@{\hspace{5mm}}}
 \multicolumn{2}{l}{Parameter} &   Value 
   \\[1.5mm]
  \hline 
  \\[-1.5mm]
$N_{\decay{\PUpsilon(2\PS)}{\PUpsilon(1\PS)\pipi}}$   & 
$\left[10^3\right]$ & 
$\phantom{10\,0}88.55\pm  1.05$ 
\\
$N_{\decay{\PUpsilon(3\PS)}{\PUpsilon(2\PS)\pipi}}$   & 
$\left[10^3\right]$ &  
$\phantom{10\,00}1.46 \pm 0.05$ 
\\ 
$m_{\PUpsilon(2\PS)}$ 
& $\left[\!\mevcc\right]$ & 
%% $10\,023.111 \pm 0.026$
$10\,023.25 \pm 0.03$
%% m_{\PUpsilon(2S)} & = & 10\,023.254 \pm 0.026
\\
$m_{\PUpsilon(3\PS)}$ 
& $\left[\!\mevcc\right]$ &
%% 10\,355.138 \pm 0.030$  
$10\,355.28 \pm 0.03$  
%% m_{\PUpsilon(3S)} & = & 10\,355.282 \pm 0.031
%%
%% \\
%% $m_{\PUpsilon(3\PS)} - m_{\PUpsilon(2\PS)}$
%% & $\left[\!\mevcc\right]$ &
%% %% $\phantom{10\,}332.028 \pm 0.040$
%% $\phantom{10\,}332.03 \pm 0.04$
\end{tabular*}
\end{table}
Since the \mbox{$\decay{\PUpsilon(1\PS)}{\mumu}$} and 
\mbox{$\decay{\PUpsilon(2\PS)}{\mumu}$}~candidates
are constrained to their known masses~\cite{PDG2024}, 
the~residual corrections to the~measured 
$\PUpsilon(2\PS)$ and $\PUpsilon(3\PS)$~masses
and their differences
are small.
Furthermore, the~corrections numerically cancel for
%% From the $\PUpsilon(2\PS)$ and $\PUpsilon(3\PS)$~mass values 
the~mass splitting %% is derived:
determined to be 
\begin{equation}
\label{eq:dmYpipi}
   m_{\PUpsilon(3\PS)}-m_{\PUpsilon(2\PS)} = 332.03 \pm 0.04 \mevcc \,,  
\end{equation}
where the uncertainty is statistical only. 

The statistical uncertainty for 
the~$\PUpsilon(2\PS)$ and 
$\PUpsilon(3\PS)$~mass difference 
%% from Table~\ref{tab:Ypipi_fits},
%% and 
from Eq.~\eqref{eq:dmYpipi} 
is slightly smaller than that 
in Table~\ref{tab:Y_fits}
despite the much lower signal yield and 
larger background level. This reflects the better mass resolution
for the~\mbox{$\decay{\PUpsilon(2\PS)}{\PUpsilon(1\PS)\pipi}$}
and \mbox{$\decay{\PUpsilon(3\PS)}{\PUpsilon(2\PS)\pipi}$}~signals.
For the mass and mass difference, 
the same sources of systematic uncertainty 
as described in Sec.~\ref{sec:mumu} are considered.  
The~dominant uncertainty is due to the~knowledge of 
the~momentum scale.
The~uncertainty related to the~signal and background parameterisation 
is estimated using a~set of alternative fit models.
%A~sum of two modified Gaussian functions,
%and a~sum of modified Gaussian and Gaussian functions 
%are used as alternative signal models.
%For alternative background description
%lower-degree polynomials are probed, as well 
%as a~product of the~three\nobreakdash-body 
%phase\nobreakdash-space function
%and the~second\nobreakdash-degree polynomial. 
%% 
The~systematic uncertainties  for the~mass measurements
from analysis of the~\mbox{$\Upsilon(1\PS)\pipi$}
and \mbox{$\Upsilon(2\PS)\pipi$}
mass spectra are summarised in Table~\ref{tab:pipi_syst}. 
The~$\TwoS$~and~$\ThreeS$~mass measurements
are made with the~$\OneS$ and $\TwoS$~masses constrained to their
known values~\cite{PDG2024}.
The~uncertainties in these values are propagated 
as an external systematic uncertainty on
the mass measurements and discussed in Sec.~\ref{sec:results}.
The~background\nobreakdash-subtracted $x_{\pipi}$~distributions
for the~\mbox{$\decay{\TwoS}{\OneS\pipi}$}
and \mbox{$\decay{\ThreeS}{\TwoS\pipi}$}~decays
are found to be similar to simulation
and previosu measurements 
by the~ARGUS~\cite{ARGUS:1986gsf}
and CLEO~\cite{CLEO:1993fsd}~collaborations.
\begin{table}[tb]
\centering
\caption{\small Systematic uncertainties
on the measurement of
the~$\Upsilon(2\PS)$ 
and $\Upsilon(3\PS)$
mass parameters, 
and the~mass difference 
from 
the~analysis 
of the~\mbox{$\decay{\PUpsilon(2\PS)}{\Upsilon(1\PS)\pipi}$}
and \mbox{$\decay{\PUpsilon(3\PS)}{\Upsilon(2\PS)\pipi}$}~decays.}
\label{tab:pipi_syst}
	\vspace{2mm}
	\begin{tabular*}{0.85\textwidth}
	{@{\hspace{5mm}}l@{\extracolsep{\fill}}ccc@{\hspace{5mm}}}
\multirow{2}{*}{Source of systematic} 
  & \multicolumn{3}{c} {Uncertainty $\left[\!\kevcc\right]$} \\
  & $m_{\TwoS}$ 
  & $m_{\ThreeS}$   
  & $m_{\PUpsilon(3\PS)}-m_{\PUpsilon(2\PS)}$ 
   \\[1.5mm]
  \hline 
  \\[-1.5mm]
Momentum scale            & 120       & 33  & 87   \\
Energy loss correction    &  20       & 20  & ---  \\
Radiative corrections     &  3        & 2   & 1    \\ 
Fit model                 &  8        & 2   & 6    
  \\[1.5mm]
  \hline 
  \\[-1.5mm]
Sum in quadrature  &  122 &  39 & 89 \\
\end{tabular*}
\end{table}

\section{$\decay{\Pchi_\bquark}{\PUpsilon(1\PS)\mumu}$~selection and mass fits}
\label{sec:Xb}

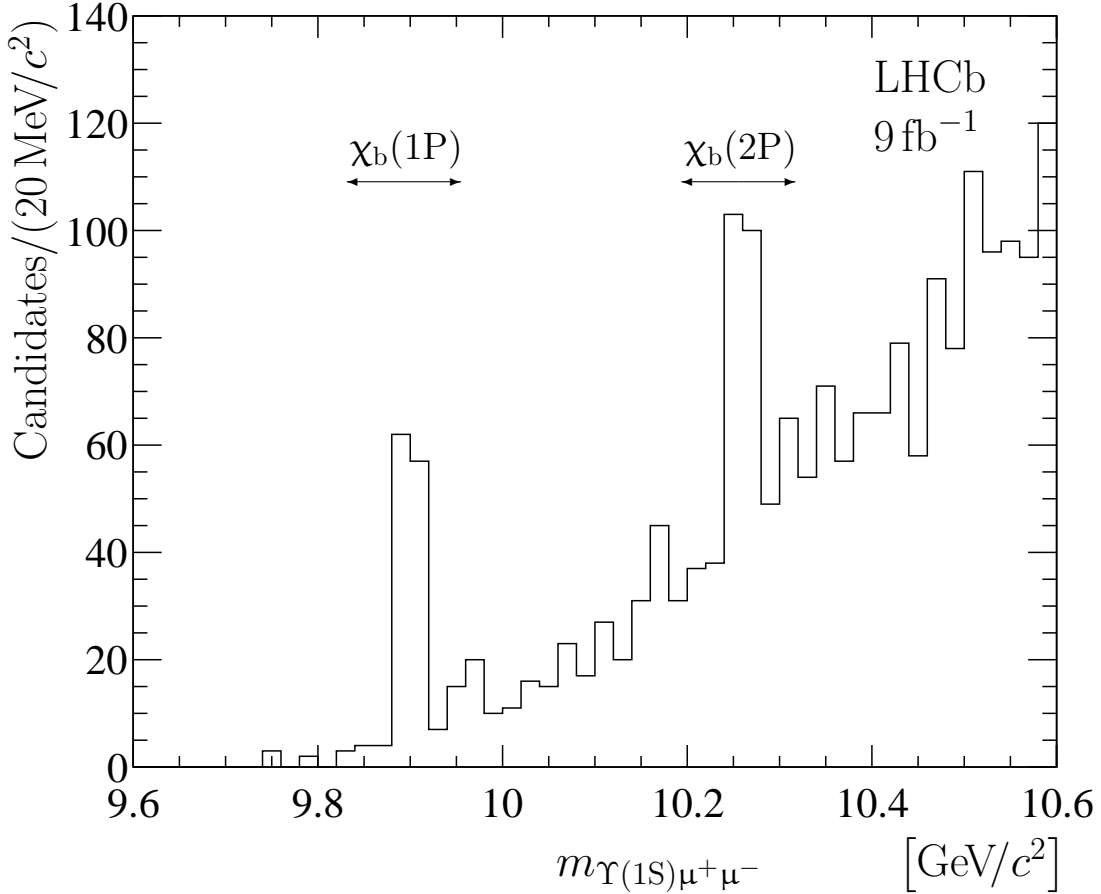
\begin{figure}[t]
  \setlength{\unitlength}{1mm}
  \centering
  \begin{picture}(150,120)
  %%
  %% \graphpaper[5](-10,-10)(170,140)
  %% 
  \put(   0,   0){ 
      \includegraphics*[width=150mm,%height=60mm,%
    %%]{Ymumu_wide.pdf}
      ]{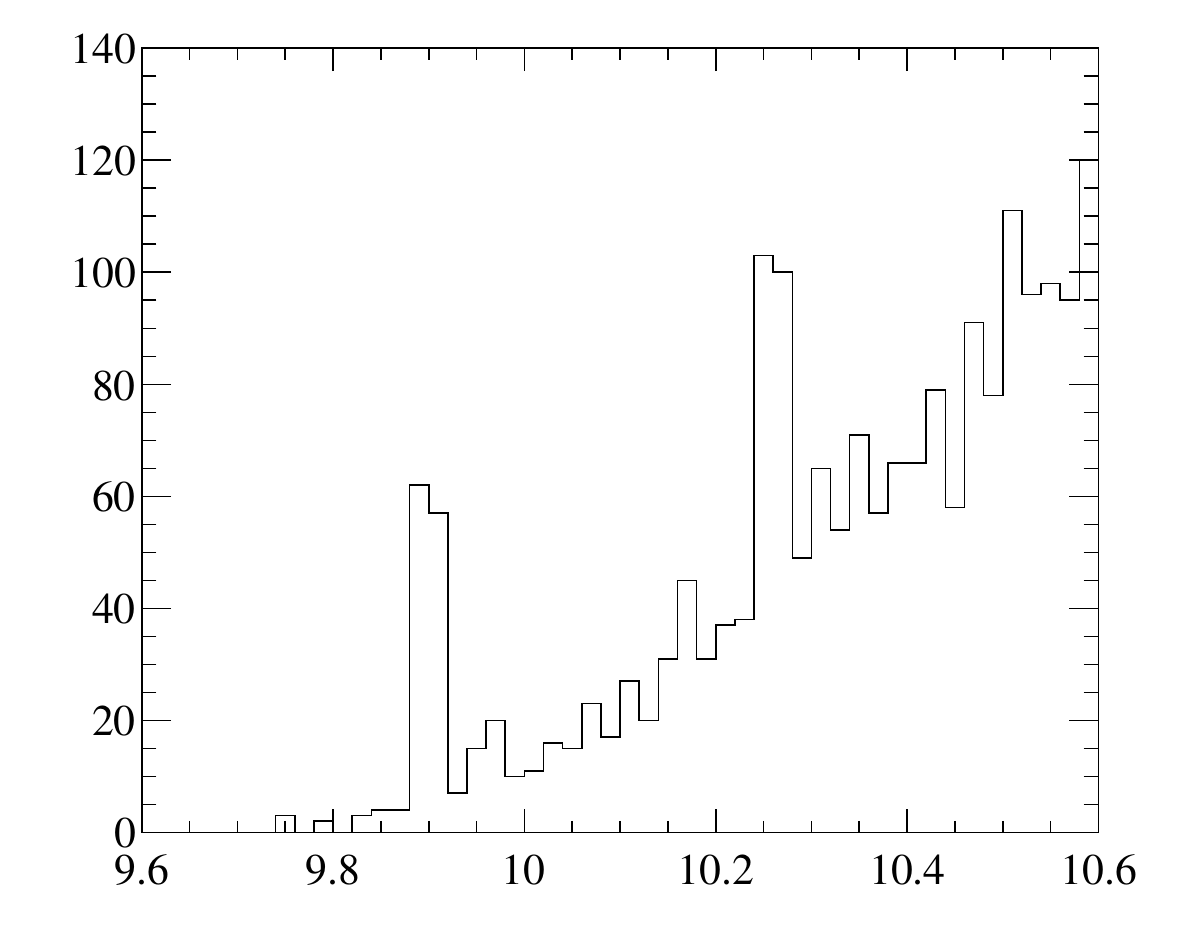}
    }
    \put(55,92){\vector( 1,0){7.5}} 
    \put(55,92){\vector(-1,0){7.5}}
    \put(48,95){\large$\Pchi_\bquark(1\PP)$}
    \put(99,92){\vector( 1,0){7.5}} 
    \put(99,92){\vector(-1,0){7.5}}
    \put(92,95){\large$\Pchi_\bquark(2\PP)$}
    \put(115,100){\Large$\begin{array}{l}\lhcb\\9\invfb\end{array}$}
    \put(75, 0){\Large$m_{\PUpsilon(1\PS)\mumu}$}  \put(120,0){\Large$\left[\!\gevcc\right]$}
    \put(3 ,55){\Large\begin{sideways}Candidates/$(20\mevcc)$\end{sideways}}
  \end{picture}
  \caption { \small
   Mass spectrum of $\PUpsilon(1\PS)\mumu$~candidates with 
   a constraint on the $\PUpsilon(1\PS)$ mass applied.
    }
  \label{fig:Ymumu_wide}
\end{figure}

Candidate $\decay{\Pchi_{\bquark}}{\PUpsilon(1\PS)\mumu}$~decays 
are formed by combining the selected 
\mbox{$\decay{\PUpsilon(1\PS)}{\mumu}$}~candidates with 
oppositely charged particles, identified as muons. In~this case,
the requirements on muon identification~\cite{LHCb-PROC-2011-008,
LHCb-DP-2012-002} 
and track quality~\cite{DeCian:2255039}
are 
sufficient to reduce the combinatorial background, mainly originating 
from pions that decay in flight. 
The only additional requirement
needed is 
on the~kinematic fit quality $\chisq_{\mathrm{fit}}/{\mathrm{ndf}}<5$. %% $\chi^2){DTF} <5$. 
The~$\PUpsilon(1\PS)\mumu$~mass spectrum,
for combinations that meet the full set of selection criteria,
is shown in Fig.~\ref{fig:Ymumu_wide}.
%% 
%% A kinematic109
%% fit [56] is made to the selected candidates, constraining the dimuon mass to the Υ(1S) or110
%% Υ(2S) mass [55], as appropriate, and requiring the candidate to be consistent with coming111
%% from a primary vertex.
 
Extended unbinned maximum-likelihood 
fits to the~$\PUpsilon(1\PS)\mumu$~mass spectrum are made 
separately for the~$\Pchi_\bquark(1\PP)$ and $\Pchi_\bquark(2\PP)$~regions, defined as 
\mbox{$9.84 < m_{\PUpsilon(1\PS)\mumu}< 9.96\gevcc$} and 
\mbox{$10.20 < m_{\PUpsilon(1\PS)\mumu}< 10.32\gevcc$}, respectively. 
For each region the~fit function consists of three components: 
two signal components, describing 
the~axial\nobreakdash-vector and tensor $\Pchi_\bquark$~states, modelled by the~modified Gaussian functions 
with shape parameters fixed from simulation,  
and a~background component 
taken as a~positive second\nobreakdash-order 
polynomial function. 
The~natural widths of the~$\Pchi_{\bquark}$ 
states are assumed to be small, 
as predicted in Refs.~\cite{Godfrey:2015dia,
Segovia:2016xqb,
Deng:2016ktl,
Wang:2018rjg,
Asghar:2023fvk}, 
and are neglected. 
In~these fits, the~width parameters 
of the~Gaussian functions
are scaled with a~factor~$s_f$,
that is Gaussian-constrained from Eq.~\eqref{eq:scale_factor}. 
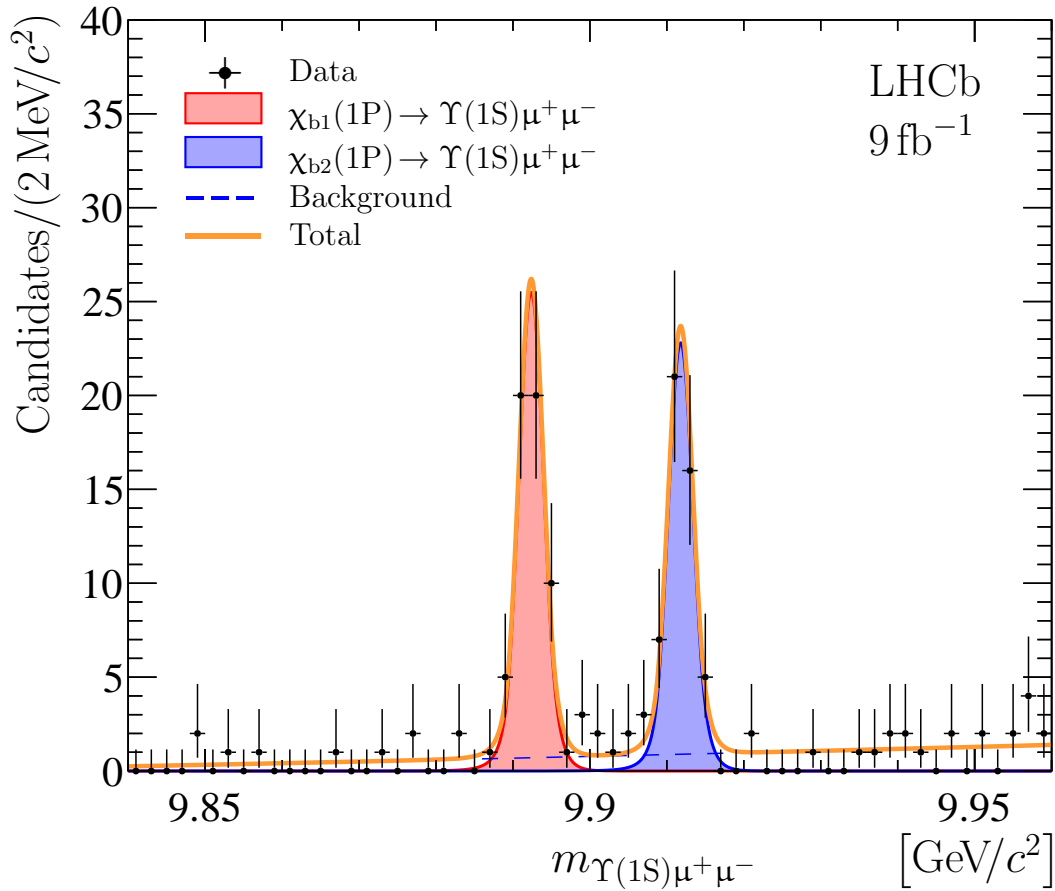
\begin{figure}[htb]
  \setlength{\unitlength}{1mm}
  \centering
  \definecolor{gr}{rgb}{0.35, 0.83, 0.33}
  \definecolor{br}{rgb}{0.43, 0.98, 0.98}
  \definecolor{vi}{rgb}{0.39, 0.37, 0.96}
  \definecolor{db}{rgb}{0.1, 0.08, 0.41}
  \begin{picture}(150,120)
  \put(   0,   0){ 
      \includegraphics*[width=150mm,%height=60mm,%
    %% ]{Xb1P.pdf}
       ]{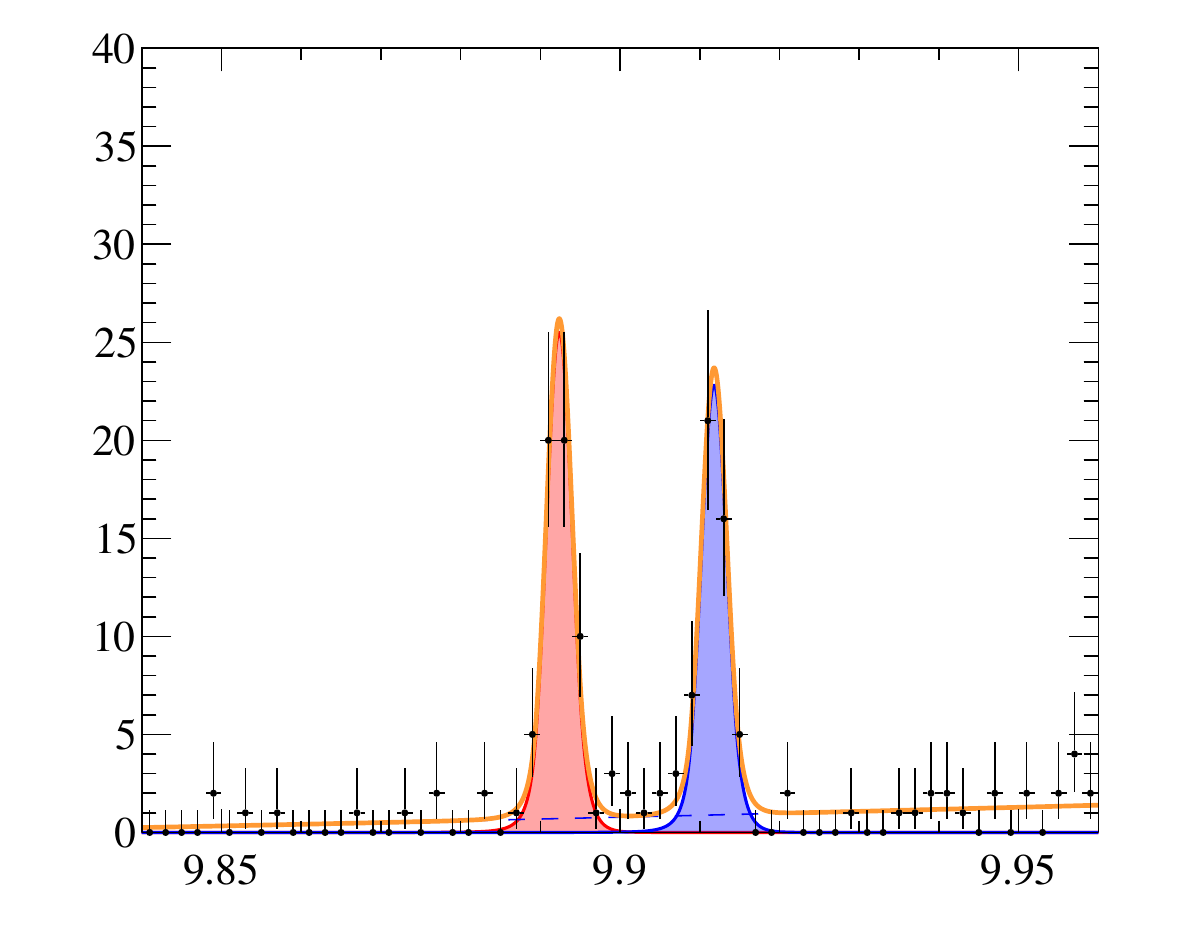}
    }
    \put(115,100){\Large$\begin{array}{l}\lhcb\\9\invfb\end{array}$}
     \put(25,95){$\begin{array}{cl}
	 %% DATA 
	 \!\!\!\bigplus\!\!\!\!\!\!\bullet\mkern-5mu&\mathrm{Data}  
	 \\ 
	 \begin{tikzpicture}[x=1mm,y=1mm]\filldraw[fill=red!35!white,draw=red,thick]  (0,0) rectangle (10,4);\end{tikzpicture} 
      & \decay{\Pchi_{\bquark1}(1\PP)}{\PUpsilon(1\PS)\mumu} 
	 \\
     \begin{tikzpicture}[x=1mm,y=1mm]\filldraw[fill=blue!35!white,draw=blue,thick]  (0,0) rectangle (10,4);\end{tikzpicture} 
      & \decay{\Pchi_{\bquark2}(1\PP)}{\PUpsilon(1\PS)\mumu} 
	 \\
      {\color[RGB]{0,0,255}{\hdashrule[0.5ex][x]{10mm}{1.3pt}{2.3mm 0.8mm}}} & \mathrm{Background}
	 \\
      {\color[RGB]{255,153,51} {\rule[0.5ex]{10mm}{2.0pt}}} & \mathrm{Total}
	 \end{array}$}
    \put(75, 1){\Large$m_{\PUpsilon(1\PS)\mumu}$}  \put(120,1){\Large$\left[\!\gevcc\right]$}
    \put( 4,59){\Large\begin{sideways}Candidates/$(2\mevcc)$\end{sideways}}
  \end{picture}
  \caption { \small
  Mass spectrum of $\PUpsilon(1\PS)\mumu$~candidates
  in the~$\Pchi_{\bquark}(1\PP)$~region
  with a constraint on the $\PUpsilon(1\PS)$ mass applied. 
  The~result of the~fit described in the~text is overlaid.
  }
  \label{fig:Xbone}
\end{figure}
\begin{figure}[htb]
  \setlength{\unitlength}{1mm}
  \centering
  \definecolor{gr}{rgb}{0.35, 0.83, 0.33}
  \definecolor{br}{rgb}{0.43, 0.98, 0.98}
  \definecolor{vi}{rgb}{0.39, 0.37, 0.96}
  \definecolor{db}{rgb}{0.1, 0.08, 0.41}
  \begin{picture}(150,120)
  \put(   0,   0){ 
      \includegraphics*[width=150mm,%height=60mm,%
    %% ]{Xb2P.pdf}
      ]{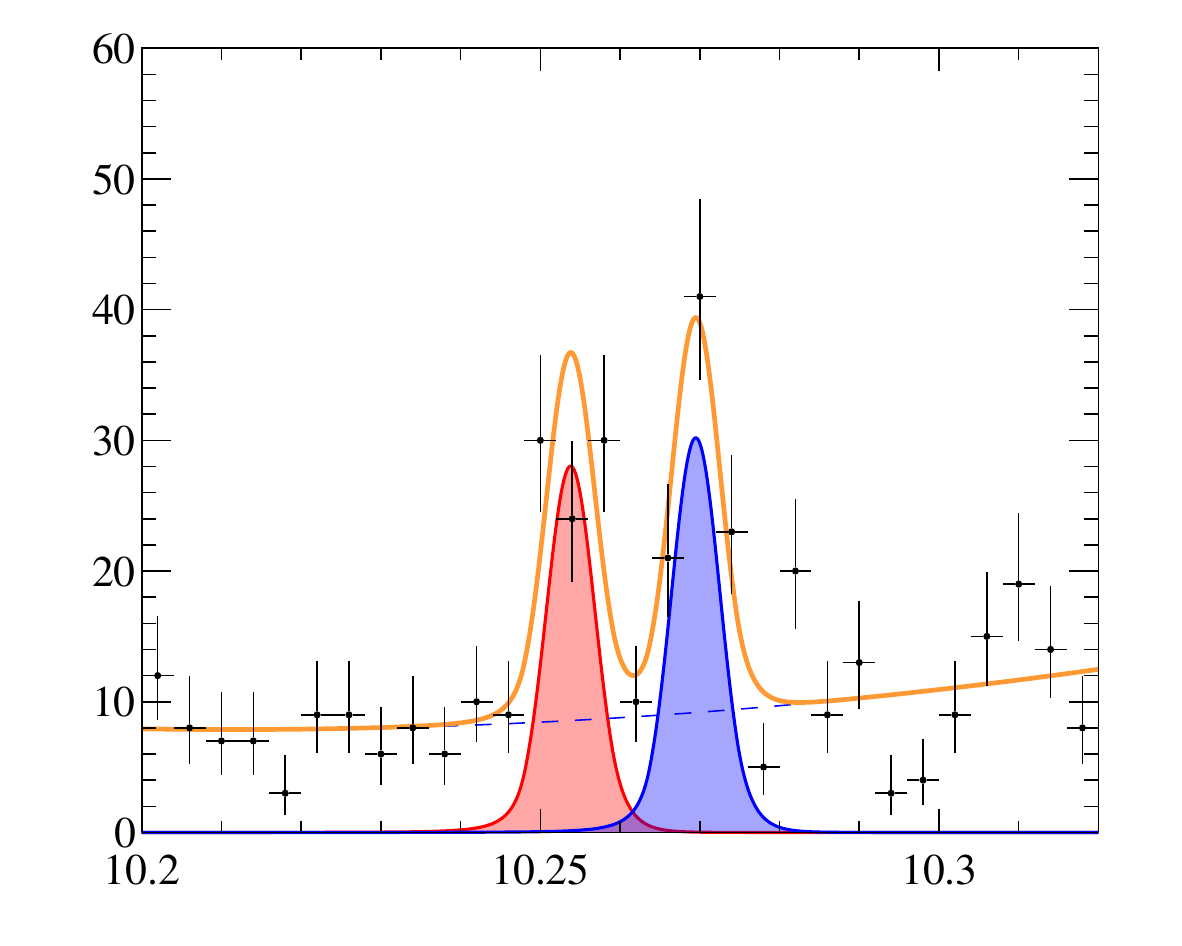}
    } 
    \put(115,100){\Large$\begin{array}{l}\lhcb\\9\invfb\end{array}$}
     \put(25,95){$\begin{array}{cl}
	 %% DATA 
	 \!\!\!\bigplus\!\!\!\!\!\!\bullet\mkern-5mu&\mathrm{Data}  
	 \\ 
	 \begin{tikzpicture}[x=1mm,y=1mm]\filldraw[fill=red!35!white,draw=red,thick]  (0,0) rectangle (10,4);\end{tikzpicture} 
      & \decay{\Pchi_{\bquark1}(2\PP)}{\PUpsilon(1\PS)\mumu} 
	 \\
     \begin{tikzpicture}[x=1mm,y=1mm]\filldraw[fill=blue!35!white,draw=blue,thick]  (0,0) rectangle (10,4);\end{tikzpicture} 
      & \decay{\Pchi_{\bquark2}(2\PP)}{\PUpsilon(1\PS)\mumu} 
	 \\
      {\color[RGB]{0,0,255}{\hdashrule[0.5ex][x]{10mm}{1.3pt}{2.3mm 0.8mm}}} & \mathrm{Background}
	 \\
      {\color[RGB]{255,153,51} {\rule[0.5ex]{10mm}{2.0pt}}} & \mathrm{Total}
	 \end{array}$}
    \put(75, 1){\Large$m_{\PUpsilon(1\PS)\mumu}$}  \put(120,1){\Large$\left[\!\gevcc\right]$}
    \put( 4,59){\Large\begin{sideways}Candidates/$(4\mevcc)$\end{sideways}}
  \end{picture}
  \caption { \small
    Mass spectrum of $\PUpsilon(1\PS)\mumu$~candidates
  in the~$\Pchi_{\bquark}(2\PP)$~region
  with a constraint on the $\PUpsilon(1\PS)$ mass applied. 
  The~result of the~fit described in the~text is overlaid.
   }
  \label{fig:Xbtwo}
\end{figure}
The~results of the~fits 
are 
overlaid in Figs.~\ref{fig:Xbone} and~\ref{fig:Xbtwo} and 
summarised in Table~\ref{tab:Ymumu_fit}.
\begin{table}[tb]
\centering
\caption{\small
The parameters of interest,
yields $N$ and masses, 
from the fits
to the~$\PUpsilon(1\PS)\mumu$~mass spectrum.
%% from Figs.~\ref{fig:Xbone} and~\ref{fig:Xbtwo}.
The~uncertainties  are statistical only.
The~last column shows 
the~statistical 
significance of the~signals 
in units of standard deviations.
    } 
	\label{tab:Ymumu_fit}
	\vspace{2mm}
	\begin{tabular*}{0.75\textwidth}
	{@{\hspace{5mm}}l@{\extracolsep{\fill}}lcl@{\hspace{5mm}}}
 \multicolumn{2}{l}{Parameter} &  \ \ Value  & $\mathscr{S}$
   \\[1.5mm]
  \hline 
  \\[-1.5mm]
$N_{\decay{\Pchi_{\bquark1}(1\PP)}{\PUpsilon(1\PS)\mumu}}$  
& &  $\phantom{10\,00}53.6 \pm 7.7\phantom{0} $  &  $12.4$
\\   
$N_{\decay{\Pchi_{\bquark2}(1\PP)}{\PUpsilon(1\PS)\mumu}}$
& &  $\phantom{10\,00}47.9 \pm 7.4\phantom{0} $  &  $11.5$
\\
$N_{\decay{\Pchi_{\bquark1}(2\PP)}{\PUpsilon(1\PS)\mumu}}$ 
& &  $\phantom{10\,00}51.1 \pm 10.4 $  & $\phantom{0}6.5$
\\
$N_{\decay{\Pchi_{\bquark2}(2\PP)}{\PUpsilon(1\PS)\mumu}}$
& &  $\phantom{10\,00}59.3 \pm 10.4 $  & $\phantom{0}7.2$
\\
$m_{\Pchi_{\bquark1}(1\PP)}$  & $\left[\!\mevcc\right]$ &  
%% 9892.33 +- 0.251442
%% $\phantom{0}9\,892.37 \pm 0.25$
$\phantom{0}9\,892.50 \pm 0.26$
\\
$m_{\Pchi_{\bquark2}(1\PP)}$  & $\left[\!\mevcc\right]$ & 
%% ( +9911.86 +- 0.287349 )
%% $\phantom{0}9\,911.39 \pm 0.29$
$\phantom{0}9\,911.92 \pm 0.29$
\\
$m_{\Pchi_{\bquark1}(2\PP)}$  & $\left[\!\mevcc\right]$ & 
%% +10254.4 +- 1.10157
%% $10\,253.80 \pm 0.75 $
$10\,253.97 \pm 0.75 $
\\
$m_{\Pchi_{\bquark2}(2\PP)}$  & $\left[\!\mevcc\right]$ & 
%% +10269.1 +- 0.734033 
%% $10\,269.49 \pm 0.67$ 
$10\,269.67 \pm 0.67$ 
\end{tabular*}
\end{table}
The~statistical 
significance 
is calculated 
for each the~four observed 
\mbox{$\decay{\Pchi_{\bquark}}{\PUpsilon(1\PS)\mumu}$}~decays 
using Wilks' 
theorem~\cite{Wilks:1938dza}
and also listed in Table~\ref{tab:Ymumu_fit}.
The mass splittings for the~$\Pchi_\bquark(1\PP)$
and $\Pchi_\bquark(2\PP)$~states are found to be 
\begin{subequations} \label{eq:chib_results_corr}
\begin{eqnarray}
\updelta m_{\Pchi_{\bquark}(1\PP)} 
& \equiv & 
m_{\Pchi_{\bquark2}(1\PP)} - m_{\Pchi_{\bquark1}(1\PP)} 
%% =  19.42 \pm 0.39 \mevcc \,,\\ 
=  19.4 \pm 0.4 \mevcc \,,\\ 
\updelta m_{\Pchi_{\bquark}(2\PP)} 
& \equiv & 
m_{\Pchi_{\bquark2}(2\PP)} - m_{\Pchi_{\bquark1}(2\PP)} 
%% =  15.70 \pm 0.98 \mevcc \,, 
=  15.7 \pm 1.0 \mevcc \,, 
\end{eqnarray}
\end{subequations}
where the uncertainties are statistical only.

Systematic uncertainties for 
the~masses of the~$\Pchi_{\bquark}$~states and mass~splittings
are evaluated using the~same techniques described in Sec.~\ref{sec:Ups_pipi} for 
the~$\PUpsilon(2\PS)$ and
$\PUpsilon(3\PS)$~masses and $\PUpsilon(3\PS)-\PUpsilon(2\PS)$~mass 
difference.
The~uncertainties  are
summarised in Table~\ref{tab:chib_syst}.
Again, the~dominant uncertainty is related to knowledge 
of the~momentum scale, and largely cancels for the~mass splittings. 
For the systematic uncertainty associated with the~fit model, 
%% the~same set of alternative signal models is probed.
a~set of alternative signal models are probed.
To~test alternative parameterisations of the~background, 
%% To~describe 
%% the~background, 
first-order polynomial
functions are used,  as well as 
a~product of a~three\nobreakdash-body phase-space function~\cite{Byckling}
and the~first- and second-order polynomials.   
An~additional systematic uncertainty is related 
to the~assumption of the~negligible natural width 
for the~$\Pchi_\bquark$~states. 
To~estimate the
associated uncertainty, 
signal shapes are parameterised 
as relativistic Breit\nobreakdash--Wigner 
functions convolved  with the~detector resolution, and 
a~series of fits are performed scanning 
the~natural widths from 50 to 250\kev, 
covering the~range of theoretical expectations~\cite{Godfrey:2015dia,
Segovia:2016xqb,
Deng:2016ktl,
Wang:2018rjg,
Asghar:2023fvk}. The~maximal deviations 
from the~results of the~baseline fits 
are taken as the~associated systematic uncertainty. 
The~mass~measurements in Table~\ref{tab:Ymumu_fit}
are made with the~$\PUpsilon(1\PS)$~mass constrained 
to the~known value~\cite{PDG2024}. 
The~uncertainty
in this value is propagated as an external systematic uncertainty on the~mass measurements.
\begin{table}[tb]
\centering
\caption{\small Systematic uncertainties
on the measurement of
the~$\Pchi_{\bquark}$~masses and mass differences. }
\label{tab:chib_syst}
	\vspace{2mm}
	\begin{tabular*}{0.99\textwidth}
	{@{\hspace{2mm}}l@{\extracolsep{\fill}}cccccc@{\hspace{2mm}}}
\multirow{2}{*}{Source of systematic} 
  & \multicolumn{6}{c} {Uncertainty $\left[\!\kevcc\right]$} \\
  & $m_{\Pchi_{\bquark1}(1\PP)}$ 
   & $m_{\Pchi_{\bquark2}(1\PP)}$ 
  & $m_{\Pchi_{\bquark1}(2\PP)}$ 
   & $m_{\Pchi_{\bquark2}(2\PP)}$ 
 & $\updelta m_{\Pchi_{\bquark}(1\PP)}$ 
   & $\updelta m_{\Pchi_{\bquark}(2\PP)}$ 
   \\[1.5mm]
  \hline 
  \\[-1.5mm]
Momentum scale              &  99     &   106  & 214 & 218 & 7   & 4   \\
Energy loss correction      &  20     &    20  &  20 &  20 & --- & --- \\
Radiative corrections       &  7      &     8  &  13 &  13 & 1   & --- \\
Fit model                   &  4      &     3  &   4 &   2 & 6   & 2   \\
Natural width               &   ---   &  ---   &  10 &  10 & --- & --- 
   \\[1.5mm]
  \hline 
  \\[-1.5mm]
Sum in quadrature  &    101  &  109    &  215  &  219   & 9 & 4 
\end{tabular*}
\end{table}
%% \todo[inline,size=large]{probably for tables
%% it is better to indicate  $<X$ instead of  $--$. What do you think? CONSULT REFEREES}
%% 
%% 
%% {\color{red}
The~statistical significance for
each $\Pchi_\bquark$~state is recalculated 
for each alternative fit model, 
and the~minimal values 
of $12.4$ , $11.2$ , $6.3$ and $7.0$~standard deviations 
are taken as the significance of the~$\Pchi_{\bquark1}(1\PP)$, 
$\Pchi_{\bquark2}(1\PP)$, 
$\Pchi_{\bquark1}(2\PP)$~and~$\Pchi_{\bquark2}(2\PP)$~states
accounting for the~systematic uncertainties.
%%}

%% \section{Summary and discussion}
\section{Results and discussion}
\label{sec:results}

A large sample of  \mbox{$\decay{\PUpsilon}{\mumu}$}~decays is used to measure the masses and mass differences of 
the~$\PUpsilon$~states.
%% While the absolute mass measurement
%% is dominated by the systematic uncertainty, 
%% the~mass differences are measured precisely 
While the~absolute mass measurements have a~large systematic uncertainty, 
the~mass differences are measured much more precisely 
\begingroup
\allowdisplaybreaks
\begin{subequations}\label{eq:res_mumu}
\begin{eqnarray}
m_{\OneS} & = & 9\,460.37 \pm 0.01  \pm 2.85 \mevcc\,, \label{eq:res_mumu_1s} \nonumber 
 \\
m_{\TwoS}  - m_{\OneS}  & = & \phantom{0\,}562.71 \pm 0.04 \pm 0.20 \mevcc \,, \label{eq:res_mumu_21}
\\
m_{\ThreeS}  - m_{\TwoS}  & = & \phantom{0\,}331.77 \pm 0.07 \pm 0.10 \mevcc \,, \label{eq:res_mumu_32}
\end{eqnarray}
\end{subequations}
\endgroup
where the first uncertainty is statistical and the second systematic. 

Using \mbox{$\decay{\PUpsilon(2\PS)}{\PUpsilon(1\PS)\pipi}$}~decays,
where the~$\OneS$~mass is constrained~\cite{PDG2024},    
a~precise measurement of the~$\PUpsilon(2\PS)$~mass is made, 
\begin{equation}
m_{\TwoS}      =  10\,023.25 \pm 0.03  \pm 0.12 \pm  0.09\mevcc \,, \label{eq:res_pipi_m2s}
\end{equation}
where the first uncertainty is statistical, the second systematic and the third from the~knowledge of 
the~$\OneS$~mass~\cite{PDG2024}.
%This is the most precise measurement of the $\TwoS$ mass to date.
%It lies between the value measured by the MD-1 collaboration~\cite{OLYA:2000toq} 
%and the reanalysis of ARGUS data by Shamov~\cite{Shamov:2022ajx}.
%% 
The uncertainty due to the~knowledge of the~$\PUpsilon(1\PS)$~mass~\cite{PDG2024} 
%% \mbox{$m_{\OneS}=9\,460.4\pm0.1\mevcc$}~\cite{PDG2024}, 
largely cancels in the~mass difference
\begin{equation}
m_{\TwoS}  - m_{\OneS}  = 562.85 \pm 0.03 \pm 0.12 \pm  0.01\mevcc\,.\label{eq:res_pipi_21} 
\end{equation}
%% \end{subequations}
This value agrees with the one obtained 
with \mbox{$\decay{\PUpsilon}{\mumu}$}~decays, Eq.~\eqref{eq:res_mumu_21},
and they are combined in Appendix~\ref{sec:dmaverage}.  %% at the $1 \sigma$ level.

Using~\mbox{$\decay{\PUpsilon(3\PS)}{\PUpsilon(2\PS)\pipi}$}~decays,
where the~$\TwoS$~mass is constrained~\cite{PDG2024}, 
a~precise measurement of the~$\PUpsilon(3\PS)$~mass 
is made, 
%% The~$\PUpsilon(3\PS)$~mass measured using 
%% the~\mbox{$\decay{\PUpsilon(3\PS)}{\PUpsilon(2\PS)\pipi}$}~decays,
%% where the~$\TwoS$~mass is constrained~\cite{PDG2024}, 
%% is
\begin{equation}
m_{\ThreeS}   =  10\,355.28 \pm 0.03 \pm 0.04 \pm 0.48\mevcc  \,, \label{eq:res_pipi_m3s_1} 
\end{equation}
where the first uncertainty is statistical, the second is systematic and the third is from the~knowledge of 
the~$\TwoS$~mass~\cite{PDG2024}, 
%%, \mbox{$m_{\TwoS}=10\,023.4\pm0.5\mevcc$}~\cite{PDG2024}. 
The~uncertainty due to the~knowledge of the~$\Upsilon(2\PS)$~mass
largely cancels in the mass difference
%rom Eqs.~\eqref{eq:res_pipi_m3s_2} and~\eqref{eq:res_pipi_m2s} 
%% \todo{what is combined here?}
\begin{equation}
m_{\ThreeS}  - m_{\TwoS}  = 331.88 \pm 0.03 \pm 0.04 \pm 0.02\mevcc \,. \label{eq:res_pipi_32_1}
\end{equation}
This is in agreement with the value obtained with 
\mbox{$\decay{\PUpsilon}{\mumu}$}~decays, Eq.~\eqref{eq:res_mumu_32}, 
and the two measurements are combined in Appendix~\ref{sec:dmaverage}. 
The~dependence on the external knowledge of the~$\TwoS$~mass is also removed by using
the~$\TwoS$~mass given in Eq.~\eqref{eq:res_pipi_m2s} instead of 
the~value given in the~2024 edition of the~particle data group averages\,(PDG'24)~\cite{PDG2024}.
Assuming the~uncertainties from
the~momentum scale and energy loss correction are fully correlated, it gives
%% \begin{equation}
%% m_{\ThreeS} = 10\,355.13 \pm 0.04  \pm  0.01 \pm 0.15 \pm  0.09\,(m_{\OneS}) \mevcc\,, \label{eq:res_pipi_m3s_2}
%% \end{equation}
%% where the first uncertainty is statistical, 
%% the second the uncorrelated systematic, 
%% the third the correlated systematic 
%% and the last due to the knowledge of the~$\OneS$~mass.
%%
\begin{equation}
m_{\ThreeS} = 10\,355.13 \pm 0.04  \pm 0.15 \pm  0.09\mevcc\,, \label{eq:res_pipi_m3s_2}
\end{equation}
where 
%% the first uncertainty is statistical, 
%% the second the uncorrelated systematic, 
%% the third the correlated systematic 
%% and 
the last uncertainty is due to the knowledge of the~$\OneS$~mass~\cite{PDG2024}.

These measurements can be compared to the current experimental knowledge 
of the~$\PUpsilon$~masses summarised 
in Tables~\ref{tab:umasspdg2} 
and~\ref{tab:umasspdg3}.
\begin{table}[tb]
\centering
\caption{\small
\label{tab:umasspdg}
%% An alternative  variant of Table~\ref{tab:umasspdg}.
Current knowledge of $\PUpsilon$ masses and mass splittings.}
\label{tab:umasspdg2}
	\vspace{2mm}
	\begin{tabular*}{0.80\textwidth}
	{@{\hspace{5mm}}l@{\extracolsep{\fill}}lcc@{\hspace{5mm}}}
 \multicolumn{2}{l}{Parameter} 
 &   PDG'22~\cite{PDG2022}
 &   PDG'24~\cite{PDG2024}
   \\[1.5mm]
  \hline 
  \\[-1.5mm]
$m_\OneS$             & $\left[\!\mevcc\right]$ &  $\phantom{1}9\,460.30 \pm 0.26$           &  $\phantom{1}9\,460.4 \pm 0.1$ 
%% & \cite{Shamov:2022ajx}\\
\\
$m_\TwoS$             & $\left[\!\mevcc\right]$ &  $10\,023.26 \pm 0.31$                     &  $10\,023.4 \pm 0.5$  
%% & \cite{Shamov:2022ajx} \\
\\ 
$m_\ThreeS$           & $\left[\!\mevcc\right]$ & $10\,355.2\phantom{0}  \pm 0.5\phantom{0}$  &  $10\,355.1  \pm 0.5$  
%% &  \cite{Shamov:2022ajx}  
\\
$m_\ThreeS - m_\TwoS$ & $\left[\!\mevcc\right]$ & 
$\phantom{10\,}331.50 \pm 0.13$  & 
$\phantom{10\,}331.50 \pm 0.13$ 
%% 
%% $m_\TwoS - m_\OneS$   & $\left[\!\mevcc\right]$ & $562.170  \pm 0.007 \pm 0.088$ &  \cite{BaBar:2011god}\\
%% $m_\ThreeS - m_\OneS$ & $\left[\!\mevcc\right]$ & $893.813  \pm 0.015\pm 0.107$ & \cite{BaBar:2011god} \\
%% $m_\ThreeS - m_\TwoS$ & $\left[\!\mevcc\right]$ & $331.50\phantom{0}  \pm 0.02\phantom{0}\pm 0.13\phantom{0}$  & \cite{BaBar:2011krt} 
\end{tabular*}
\end{table}
\begin{table}[t]
\centering
\caption{\small
The $\PUpsilon$ mass splittings measured by the~BaBar collaboration~\cite{BaBar:2011god}. %% ,BaBar:2011krt}.
These~results are not considered in any edition of the PDG. 
}
\label{tab:umasspdg3}
	\vspace{2mm}
	\begin{tabular*}{0.65\textwidth}
	{@{\hspace{5mm}}l@{\extracolsep{\fill}}lc@{\hspace{5mm}}}
 \multicolumn{2}{l}{Parameter} 
 &   Value
 %% &   Ref.
   \\[1.5mm]
  \hline 
  \\[-1.5mm]
$m_\TwoS - m_\OneS$   & $\left[\!\mevcc\right]$ & $562.170  \pm 0.007 \pm 0.088$ %% &  \cite{BaBar:2011god}
\\ 
$m_\ThreeS - m_\OneS$ & $\left[\!\mevcc\right]$ & $893.813  \pm 0.015\pm 0.107$  %% & \cite{BaBar:2011god} %%  \\
%% $m_\ThreeS - m_\TwoS$ & $\left[\!\mevcc\right]$ & $331.50\phantom{0}  \pm 0.02\phantom{0}\pm 0.13\phantom{0}$  & \cite{BaBar:2011krt} 
\end{tabular*}
\end{table}
Direct measurements of the $\PUpsilon$ masses come from studies made in 
the~1980s by the~MD\nobreakdash-1~collaboration
at the~VEPP\nobreakdash-4~storage rings~\mbox{\cite{Artamonov:1983vz,Baru:1986mi,Baru:1992jnf}}, 
the~CUSB collaboration
at the~CESR~accelerator~\cite{MacKay:1984kv}, 
and a~joint study by the~ARGUS and Crystal Ball collaborations
at the~DORIS~II~collider~\cite{ARGUS:1983mhk}. 
Recently, these data have been reanalysed~\cite{Shamov:2022ajx} 
to take into account improved knowledge of radiative corrections~\cite{Kuraev:1985hb},  
interference effects~\cite{Azimov:1975ft,Anashin:2011ku}
and the~electron mass shift~\cite{Cohen:1987fr}.
The~reanalysis resolves long-standing tensions between the CUSB and the MD-1 collaboration results, 
improving the~knowledge of the~$\OneS$ mass significantly in the PDG averages.
Prior to PDG'24, the~$\TwoS$~mass average 
included the~data from the ARGUS and Crystal~Ball collaborations~\cite{ARGUS:1983mhk}. 
Reference~\cite{Shamov:2022ajx} has also studied these data,  changing the~central value
by~$0.15 \mevcc$. 
Neither the~value given in Ref.~\cite{Shamov:2022ajx} nor 
the~original value obtained in 
Ref.~\cite{ARGUS:1983mhk}
are included in the~PDG'24 averages. 
Consequently, the~uncertainty on the~$\TwoS$~mass has worsened 
from $0.3$~to~$0.5\mevcc$.  
The~values of the~$\TwoS$ and $\ThreeS$~mass presented here agree 
well with those from Ref.~\cite{Shamov:2022ajx}. 
The~value of $m_{\TwoS}$ also agrees with the~value~\cite{Shamov:2022ajx} 
obtained from the~reanalysis of ARGUS and Crystal Ball data, 
$m_{\TwoS} = 10\,022.7 \pm  0.4 \mevcc$. 
For~the~mass differences, there are tensions in the results presented 
here compared to those of the~BaBar collaboration~\cite{BaBar:2011god,BaBar:2011krt}
at the level of $2-4 \sigma$.\footnote{The measurements in Ref.~\cite{BaBar:2011god} are not listed in the PDG averages.}
%% \todo{should it be 69 instead of 80?}

\begin{table}[t]
\centering
\caption{\small
 %% Knowledge of the $\Pchi_\bquark$ masses and mass differences from PDG'24~\cite{PDG2024}. 
 Masses and mass differences for the $\Pchi_\bquark$~states from PDG'24~\cite{PDG2024}. 
 For~the~$\Pchi_\bquark(1\PP)$ 
 and $\Pchi_\bquark(2\PP)$~states, 
 the second uncertainty comes from the uncertainty on 
 the~$\PUpsilon(2\PS)$ 
 and $\PUpsilon(3\PS)$~mass, respectively.  
 The last column shows
 the values
 recalculated  
 from the~photon energy in 
 the~\mbox{$\decay{\PUpsilon(2\PS)}{\Pchi_\bquark(1\PP)\g}$}~decays 
 using the~mass of~\TwoS~meson
 from PDG'24~\cite{PDG2024}
 and their difference. 
 This~calculation ignores any correlation between the photon energy measurements.
 } 
 \label{tab:chibmasspdg2}
%% \begin{scriptsize}
\label{tab:chibpdg}
	\vspace{2mm}
	\begin{tabular*}{0.85\textwidth}
	{@{\hspace{5mm}}l@{\extracolsep{\fill}}lcc@{\hspace{5mm}}}
 \multicolumn{2}{l}{Parameter} 
 &   PDG'24~\cite{PDG2024}
 &   Recalculated  
   \\[1.5mm]
  \hline 
  \\[-1.5mm]
$m_{\Pchi_{\bquark1}(1\PP)} $ & $\left[\!\mevcc\right]$ 
& $\phantom{1}9\,892.78 \pm 0.26 \pm 0.31$  
& $\phantom{1}9\,892.92 \pm 0.33 \pm 0.50$
\\
$m_{\Pchi_{\bquark2}(1\PP)}$ & $\left[\!\mevcc\right]$  
& $\phantom{1}9\,912.21 \pm 0.26 \pm 0.31$  
& $\phantom{1}9\,912.35 \pm 0.29 \pm 0.50$   
\\
$m_{\Pchi_{\bquark1}(2\PP)}$ & $\left[\!\mevcc\right]$  
& $10\,255.46 \pm  0.22 \pm 0.50$ 
\\
$m_{\Pchi_{\bquark2}(2\PP)}$ & $\left[\!\mevcc\right]$  
& $10\,268.65 \pm 0.22 \pm 0.50$  
\\
$\updelta m_{\Pchi_{\bquark}(1\PP)}$ & $\left[\!\mevcc\right]$  
& $\phantom{10\,0}19.10 \pm 0.25 \phantom{\pm 0.00}\,\,\,$ 
%% & 
& $\phantom{10\,0}19.42 \pm 0.44 \phantom{\pm 0.00}\,\,\,$
\\
$\updelta m_{\Pchi_{\bquark}(2\PP)}$ & $\left[\!\mevcc\right]$ 
& $\phantom{10\,0}13.10 \pm 0.24 \phantom{\pm 0.00}\,\,\,$  
& 
\end{tabular*}
\end{table}

The~first observation of the~\mbox{$\decay{\Pchi_{\bquark}}{\PUpsilon(1\PS)\mumu}$}~decays
provides measurements of the~masses of the~$\Pchi_\bquark$ states,
%% and the corresponding splittings: 
%% \begin{subequations} \label{eq:chib_results}
\begingroup
\allowdisplaybreaks
\begin{eqnarray*}
  m_{\Pchi_{\bquark1}(1\PP)}       & = &  \phantom{0}9\,892.50 \pm 0.26 \pm 0.10 \pm 0.10 \mevcc \,, \\ 
  m_{\Pchi_{\bquark2}(1\PP)}       & = &  \phantom{0}9\,911.92 \pm 0.29  \pm 0.11 \pm 0.10 \mevcc \,, \\ 
  m_{\Pchi_{\bquark1}(2\PP)}       & = &  10\,253.97 \pm 0.75  \pm 0.22 \pm 0.09 \mevcc \,, \\ 
  m_{\Pchi_{\bquark2}(2\PP)}       & = &  10\,269.67 \pm 0.67  \pm  0.22\pm 0.09  \mevcc \,,
\end{eqnarray*}
\endgroup
where the first uncertainty is statistical, 
the second systematic and the third from the~knowledge of the~$\OneS$ mass~\cite{PDG2024}. 
The corresponding mass splittings are
\begingroup
\allowdisplaybreaks
\begin{eqnarray*}
%% \updelta m_{\Pchi_{\bquark}(1\PP)} & = &  \phantom{10\,0}19.42 \pm 0.39 \pm 0.009 \mevcc \,, \\ 
%% \updelta m_{\Pchi_{\bquark}(2\PP)} & = &  \phantom{10\,0}15.70 \pm 0.98 \pm 0.004 \mevcc \,,
%% \updelta m_{\Pchi_{\bquark}(1\PP)} & = &  \phantom{10\,0}19.4\phantom{0} \pm 0.4 \mevcc \,, \\ 
%% \updelta m_{\Pchi_{\bquark}(2\PP)} & = &  \phantom{10\,0}15.7\phantom{0} \pm 1.0 \mevcc \,,
m_{\Pchi_{\bquark2}(1\PP)} - 
m_{\Pchi_{\bquark1}(1\PP)} 
& = &  19.4 \pm 0.4 \mevcc \,, \\  
m_{\Pchi_{\bquark2}(2\PP)} - 
m_{\Pchi_{\bquark1}(2\PP)}
& = &  15.7 \pm 1.0 \mevcc \,.
\end{eqnarray*}
\endgroup
%% \end{subequations} 
%% \todo[inline]{How to round the uncertainties for the ast two lines?}
%% where the first uncertainty is statistical, 
%% the second systematic and the third from the~knowledge of the~$\OneS$ mass. 
The~systematic uncertainties for 
the~mass splittings are negligible with respect 
to the~statistical uncertainties, and are omitted. 
These measurements can be compared to the~corresponding world averages, 
given 
%% in the 2024 edition of the 
PDG'24~\cite{PDG2024} and shown in Table~\ref{tab:chibmasspdg2}. 
The~previous measurements were largely made using the photon energy measured 
in feed down transitions, such as \mbox{$\decay{\TwoS}{\Pchi_\bquark(1\PP)\g}$}. 
%% In~the 2024 edition of the PDG 
In~the~PDG'24 
these values have not been updated to be 
consistent with the~quoted value of the~$\TwoS$~mass, which has
a~changed central value and worse precision. The third column
in Table~\ref{tab:chibmasspdg2} shows the precision of the~$\Pchi_\bquark(1\PP)$~masses
%% new value they quote 
%As this discussed above,  the central value would change and the precision 
worsening, as discussed above.

The values obtained by this analysis are the most precise measurements 
of the~$\Pchi_\bquark(1\PP)$~masses to date --
regardless of the~inconsistency in the~uncertainties quoted in PDG'24.
They~agree well with the~previous measurements, made using the~photon energy in 
\mbox{$\decay{\TwoS}{\Pchi_\bquark(1\PP)\g}$}~radiative transitions, which are 
dominated by results from the CLEO collaboration~\cite{CLEO:2004jkt} and
have very different systematic uncertainties. 
The~measurements of the~$\Pchi_\bquark(2\PP)$~masses have slightly worse precision
than the PDG'24 values. Again, the~systematic uncertainties are very different 
to previous measurements using the photon energy in 
\mbox{$\decay{\ThreeS}{\Pchi_\bquark(2\PP) \g}$}~radiative transitions. 
All the mass measurements agree well with the PDG'24 values. 
The~uncertainties on the~$\Pchi_\bquark$~mass splittings 
are not yet competitive with world averages values, 
which are dominated by high-precision measurements from 
the~BaBar collaboration~\cite{Lees:2014qea}. 
The~central value for~$\updelta m_{\Pchi_{\bquark}(1\PP)}$ 
is in good agreement with the~PDG'24 value, 
with a~precision factor of 1.6~worse. 
The~precision for $\updelta m_{\Pchi_{\bquark}(2\PP)}$ splitting 
is about four times worse than the~PDG'24 value. 
The~central value agrees with the~PDG'24 at the~level 
of $2.6$~standard deviations. 
%% \footnote{Our $\Pchi_{\bquark2}(2\PP)$ mass is $1\sigma$ higher than the PDG, 
%% the $\Pchi_{\bquark1}(2\PP)$ is $1 \sigma$ lower.}. 
%%
%% Since the~LHCb measurements are dominated by the~statistical uncertainty, 
%% they will be improved with the~larger dataset collected with 
%% the~upgraded LHCb detector~\cite{LHCb-DP-2022-002,LHCb-TDR-023}.
%%
The~mass differences between 
the~$\Pchi_{\bquark}$~and $\OneS$~states are listed
in Appendix~\ref{sec:dmxb}. 

In summary, precise spectroscopy of the hidden-beauty system is reported.
Precision measurements of the masses of the~$\TwoS$  and $\ThreeS$ states 
have been made using their decays to \mbox{$\OneS\pipi$} and \mbox{$\TwoS\pipi$}~final states, respectively. 
These measurements are competitive and are in agreement with the world averages. 
They improve the~knowledge of these parameters significantly and with different systematic 
uncertainties with respect to measurements at $\epem$~colliders.
In~addition, the~\mbox{$\OneS\mumu$} system at low mass has been studied for the first time, complementing the LHCb search for tetraquarks at higher mass reported in Ref.~\cite{LHCb-PAPER-2018-027}. 
This has allowed the~$\Pchi_\bquark(1\PP)$ and $\Pchi_{\bquark}(2\PP)$~decay 
modes to be observed with high significance, leading 
to competitive measurements of the masses of these states. 
The latter measurements are statistically limited and can be improved using the larger dataset
that will be collected with the~upgraded LHCb detector~\cite{LHCb-DP-2022-002,LHCb-TDR-023}.
With a~larger dataset, the~natural widths of these states could 
also be probed, and it will be possible to study $\Pchi_\bquark(3\PP)$~decays to 
the \mbox{$\OneS\mumu$} and \mbox{$\TwoS\mumu$}~final states.

% Do not include this in any draft (just for information in the template)
%%\input{acknowledgements_intro}
% Comment this in for paper drafts; do not include this in analysis note, conference and figure reports

\section*{Acknowledgements}
%
% These Acknowledgements valid from 3-May-2019
%
\noindent We express our gratitude to our colleagues in the CERN
accelerator departments for the~excellent performance of the LHC. 
We~thank the technical and administrative staff at the LHCb
institutes.
We acknowledge support from CERN and from the national agencies:
CAPES, CNPq, FAPERJ and FINEP\,(Brazil); 
MOST and NSFC\,(China); 
CNRS/IN2P3\,(France); 
BMBF, DFG and MPG\,(Germany); 
INFN\,(Italy); 
NWO\,(Netherlands); 
MNiSW and NCN\,(Poland); 
MCID/IFA (Romania); 
%MSHE (Russia); 
MICIU and AEI\,(Spain);
SNSF and SER\,(Switzerland); 
NASU\,(Ukraine); 
STFC\,(United Kingdom); 
DOE NP and NSF\,(USA).
We~acknowledge the computing resources that are provided by CERN, 
IN2P3\,(France), 
KIT and DESY\,(Germany),
INFN\,(Italy), 
SURF\,(Netherlands),
PIC\,(Spain), 
GridPP\,(United Kingdom), 
%RRCKI and Yandex LLC (Russia), 
CSCS\,(Switzerland), 
IFIN\nobreakdash-HH\,(Romania), 
CBPF\,(Brazil),
and Polish WLCG\,(Poland).
We are indebted to the communities behind the multiple open-source
software packages on which we depend.
Individual groups or members have received support from
ARC and ARDC\,(Australia);
Key Research Program of Frontier Sciences of CAS, CAS PIFI, CAS CCEPP, 
Fundamental Research Funds for the Central Universities, 
and Sci. \& Tech. Program of Guangzhou\,(China);
Minciencias\,(Colombia);
EPLANET, Marie Sk\l{}odowska\nobreakdash-Curie Actions, 
ERC and NextGenerationEU\,(European Union);
A*MIDEX, ANR, IPhU and Labex P2IO, and R\'{e}gion Auvergne\nobreakdash-Rh\^{o}ne\nobreakdash-Alpes\,(France);
%RFBR, RSF and Yandex LLC (Russia);
AvH Foundation\,(Germany);
ICSC\,(Italy); 
%GVA, XuntaGal, GENCAT, Inditex, InTalent and Prog.~Atracci\'on Talento, CM (Spain);
Severo Ochoa and Mar\'ia de Maeztu Units of Excellence, GVA, XuntaGal, GENCAT,
InTalent-Inditex and Prog. ~Atracci\'on Talento CM (Spain);
SRC\,(Sweden);
the~Leverhulme Trust, the~Royal Society
 and UKRI\,(United Kingdom).

%% APPENDICES 

\appendix 
%% renewcommand{\thetable}{S\arabic{table}}   
%% \renewcommand{\thefigure}{S\arabic{figure}}
%% \renewcommand{\theequation}{S\arabic{equation}}
%% \setcounter{figure}{0}
%% \setcounter{table}{0}
\setcounter{equation}{0}
\renewcommand{\theequation}{\thesection\arabic{equation}}

\section{Combination of results in the dimuon and dipion channels}
\label{sec:dmaverage}
The measurements of the mass splittings between the~$\PUpsilon$~states
using the~\mbox{$\decay{\PUpsilon}{\mumu}$}~decays, Eqs.~\eqref{eq:res_mumu},
and 
%% using the~\mbox{$\decay{\PUpsilon}{\PUpsilon\pipi}$}~decays, 
using 
the~\mbox{$\decay{\TwoS}{\OneS\pipi}$}
and 
\mbox{$\decay{\ThreeS}{\TwoS\pipi}$}
decays,
Eqs.~\eqref{eq:res_pipi_21}  
and~\eqref{eq:res_pipi_32_1}, 
are in good agreement. 
They~are combined assuming the uncertainty due to the momentum scale is fully correlated. 
%% These combinations are done assuming that the uncertainty due to the momentum scale is fully correlated. 
This gives
\begin{subequations}
%% \begin{equation}
\begin{eqnarray}
%% m_{\TwoS}  - m_{\OneS}  & =  & 562.84 \pm \, 0.02  \pm 0.03 \syst \pm \, 0.13  \mevcc \,, \label{eq:avg12massdiff}
m_{\TwoS}  - m_{\OneS}  & =  & 562.84 \pm  0.02  \pm 0.13  \mevcc \,, \label{eq:avg12massdiff}
%% \end{equation}
%% and 
%% \begin{equation}
\\ 
%% m_{\ThreeS}  - m_{\TwoS}  & =  & 331.86 \pm \, 0.03  \pm 0.02 \syst \pm \, 0.05  \mevcc \,.  \label{eq:avg32massdiff}
m_{\ThreeS}  - m_{\TwoS}  & =  & 331.86 \pm 0.03 \pm  0.05  \mevcc \,.  \label{eq:avg32massdiff}
%% \end{equation}
\end{eqnarray}
\end{subequations}
%% where the first uncertainty is statistical, the second due to the uncorrelated part of the systematic uncertainty and the third due to the knowledge of the momentum scale.
The~dipion modes dominate the determination
of the averages.

\section{Mass differences for the~$\Pchi_\bquark$ and $\PUpsilon(1\PS)$~states}
\label{sec:dmxb}
The~$\Pchi_\bquark$~mass measurements, made using the~known value of the~mass of the~$\PUpsilon(1\PS)$~meson,
are equivalent to the measurement of the~mass~differences: 
\begingroup
\allowdisplaybreaks
\begin{subequations}
\begin{eqnarray}
  m_{\Pchi_{\bquark1}(1\PP)} - m_{\PUpsilon(1\PS)} & = &   432.10 \pm 0.26 \pm 0.10   \mevcc \,, \\ 
  m_{\Pchi_{\bquark2}(1\PP)} - m_{\PUpsilon(1\PS)} & = &   451.52 \pm 0.29 \pm 0.11   \mevcc \,, \\ 
  m_{\Pchi_{\bquark1}(2\PP)} - m_{\PUpsilon(1\PS)} & = &   793.57 \pm 0.75 \pm 0.22 \mevcc \,, \\ 
  m_{\Pchi_{\bquark2}(2\PP)} - m_{\PUpsilon(1\PS)} & = &   809.27 \pm 0.67 \pm 0.22  \mevcc \,.
\end{eqnarray}
\end{subequations}
\endgroup 
%% {\color{red}
While the systematic 
uncertainties on the $\Pchi_{b}$ masses in Table~\ref{tab:chib_syst}
are dominated by the~momentum scale and hence highly correlated, the~statistical correlations 
are small, 
less than 1\% for the 
$m_{\Pchi_{\bquark1}(1\PP)}$ 
and  $m_{\Pchi_{\bquark2}(1\PP)}$~states, 
and 6\% for the 
$m_{\Pchi_{\bquark1}(2\PP)}$ 
and  $m_{\Pchi_{\bquark2}(2\PP)}$~states.
%% }

%% \input{appendices}

%\input{appendix}

% This should be taken out in the final paper

%% \clearpage
%% \input{supplementary-app}

\clearpage 

\addcontentsline{toc}{section}{References}
%\setboolean{inbibliography}{true}
\bibliographystyle{LHCb}
\bibliography{main,standard,LHCb-PAPER,LHCb-CONF,LHCb-DP,LHCb-TDR}

\ifx\mcitethebibliography\mciteundefinedmacro
\PackageError{LHCb.bst}{mciteplus.sty has not been loaded}
{This bibstyle requires the use of the mciteplus package.}\fi
\providecommand{\href}[2]{#2}
\begin{mcitethebibliography}{10}
\mciteSetBstSublistMode{n}
\mciteSetBstMaxWidthForm{subitem}{\alph{mcitesubitemcount})}
\mciteSetBstSublistLabelBeginEnd{\mcitemaxwidthsubitemform\space}
{\relax}{\relax}

\bibitem{Eichten:1978tg}
E.~Eichten {\em et~al.},
  \ifthenelse{\boolean{articletitles}}{\emph{{Charmonium: The model}},
  }{}\href{https://doi.org/10.1103/PhysRevD.17.3090}{Phys.\ Rev.\  \textbf{D17}
  (1978) 3090}, Erratum \href{https://doi.org/10.1103/physrevd.21.313.2}{ibid.\
    \textbf{D21} (1980) 313}\relax
\mciteBstWouldAddEndPuncttrue
\mciteSetBstMidEndSepPunct{\mcitedefaultmidpunct}
{\mcitedefaultendpunct}{\mcitedefaultseppunct}\relax
\EndOfBibitem
\bibitem{Godfrey:2015dia}
S.~Godfrey and K.~Moats,
  \ifthenelse{\boolean{articletitles}}{\emph{{Bottomonium mesons and strategies
  for their observation}},
  }{}\href{https://doi.org/10.1103/PhysRevD.92.054034}{Phys.\ Rev.\
  \textbf{D92} (2015) 054034},
  \href{http://arxiv.org/abs/1507.00024}{{\normalfont\ttfamily
  arXiv:1507.00024}}\relax
\mciteBstWouldAddEndPuncttrue
\mciteSetBstMidEndSepPunct{\mcitedefaultmidpunct}
{\mcitedefaultendpunct}{\mcitedefaultseppunct}\relax
\EndOfBibitem
\bibitem{Ferretti:2013vua}
J.~Ferretti and E.~Santopinto,
  \ifthenelse{\boolean{articletitles}}{\emph{{Higher mass bottomonia}},
  }{}\href{https://doi.org/10.1103/PhysRevD.90.094022}{Phys.\ Rev.\
  \textbf{D90} (2014) 094022},
  \href{http://arxiv.org/abs/1306.2874}{{\normalfont\ttfamily
  arXiv:1306.2874}}\relax
\mciteBstWouldAddEndPuncttrue
\mciteSetBstMidEndSepPunct{\mcitedefaultmidpunct}
{\mcitedefaultendpunct}{\mcitedefaultseppunct}\relax
\EndOfBibitem
\bibitem{Wang:2018rjg}
J.-Z. Wang, Z.-F. Sun, X.~Liu, and T.~Matsuki,
  \ifthenelse{\boolean{articletitles}}{\emph{{Higher bottomonium zoo}},
  }{}\href{https://doi.org/10.1140/epjc/s10052-018-6372-1}{Eur.\ Phys.\ J.\
  \textbf{C78} (2018) 915},
  \href{http://arxiv.org/abs/1802.04938}{{\normalfont\ttfamily
  arXiv:1802.04938}}\relax
\mciteBstWouldAddEndPuncttrue
\mciteSetBstMidEndSepPunct{\mcitedefaultmidpunct}
{\mcitedefaultendpunct}{\mcitedefaultseppunct}\relax
\EndOfBibitem
\bibitem{Brambilla:2010cs}
N.~Brambilla {\em et~al.}, \ifthenelse{\boolean{articletitles}}{\emph{{Heavy
  quarkonium: Progress, puzzles, and opportunities}},
  }{}\href{https://doi.org/10.1140/epjc/s10052-010-1534-9}{Eur.\ Phys.\ J.\
  \textbf{C71} (2011) 1534},
  \href{http://arxiv.org/abs/1010.5827}{{\normalfont\ttfamily
  arXiv:1010.5827}}\relax
\mciteBstWouldAddEndPuncttrue
\mciteSetBstMidEndSepPunct{\mcitedefaultmidpunct}
{\mcitedefaultendpunct}{\mcitedefaultseppunct}\relax
\EndOfBibitem
\bibitem{Gross:2022hyw}
F.~Gross {\em et~al.}, \ifthenelse{\boolean{articletitles}}{\emph{{50 Years of
  Quantum Chromodynamics}},
  }{}\href{https://doi.org/10.1140/epjc/s10052-023-11949-2}{Eur.\ Phys.\ J.\
  \textbf{C83} (2023) 1125},
  \href{http://arxiv.org/abs/2212.11107}{{\normalfont\ttfamily
  arXiv:2212.11107}}\relax
\mciteBstWouldAddEndPuncttrue
\mciteSetBstMidEndSepPunct{\mcitedefaultmidpunct}
{\mcitedefaultendpunct}{\mcitedefaultseppunct}\relax
\EndOfBibitem
\bibitem{LHCb-PAPER-2017-036}
LHCb collaboration, R.~Aaij {\em et~al.},
  \ifthenelse{\boolean{articletitles}}{\emph{{$\Pchi_{\cquark1}$ and
  $\Pchi_{\cquark2}$ resonance parameters with the~decays
  \mbox{\decay{\Pchi_{\cquark1,\cquark2}}{\jpsi\mumu}}}},
  }{}\href{https://doi.org/10.1103/PhysRevLett.119.221801}{Phys.\ Rev.\ Lett.\
  \textbf{119} (2017) 221801},
  \href{http://arxiv.org/abs/1709.04247}{{\normalfont\ttfamily
  arXiv:1709.04247}}\relax
\mciteBstWouldAddEndPuncttrue
\mciteSetBstMidEndSepPunct{\mcitedefaultmidpunct}
{\mcitedefaultendpunct}{\mcitedefaultseppunct}\relax
\EndOfBibitem
\bibitem{Lees:2014qea}
BaBar collaboration, J.~P. Lees {\em et~al.},
  \ifthenelse{\boolean{articletitles}}{\emph{{Bottomonium spectroscopy and
  radiative transitions involving the $\Pchi_{\bquark\PJ}(1\PP,2\PP)$ states
  at~\babar}}, }{}\href{https://doi.org/10.1103/PhysRevD.90.112010}{Phys.\
  Rev.\  \textbf{D90} (2014) 112010},
  \href{http://arxiv.org/abs/1410.3902}{{\normalfont\ttfamily
  arXiv:1410.3902}}\relax
\mciteBstWouldAddEndPuncttrue
\mciteSetBstMidEndSepPunct{\mcitedefaultmidpunct}
{\mcitedefaultendpunct}{\mcitedefaultseppunct}\relax
\EndOfBibitem
\bibitem{CLEO:2004jkt}
CLEO collaboration, M.~Artuso {\em et~al.},
  \ifthenelse{\boolean{articletitles}}{\emph{{Photon transitions in $\TwoS$ and
  $\ThreeS$ decays}},
  }{}\href{https://doi.org/10.1103/PhysRevLett.94.032001}{Phys.\ Rev.\ Lett.\
  \textbf{94} (2005) 032001},
  \href{http://arxiv.org/abs/hep-ex/0411068}{{\normalfont\ttfamily
  arXiv:hep-ex/0411068}}\relax
\mciteBstWouldAddEndPuncttrue
\mciteSetBstMidEndSepPunct{\mcitedefaultmidpunct}
{\mcitedefaultendpunct}{\mcitedefaultseppunct}\relax
\EndOfBibitem
\bibitem{CrystalBall:1985pow}
Crystal Ball collaboration, W.~S. Walk {\em et~al.},
  \ifthenelse{\boolean{articletitles}}{\emph{{The~$\Pchi_\bquark$~states in
  exclusive radiative decay of the~$\TwoS$}},
  }{}\href{https://doi.org/10.1103/PhysRevD.34.2611}{Phys.\ Rev.\  \textbf{D34}
  (1986) 2611}\relax
\mciteBstWouldAddEndPuncttrue
\mciteSetBstMidEndSepPunct{\mcitedefaultmidpunct}
{\mcitedefaultendpunct}{\mcitedefaultseppunct}\relax
\EndOfBibitem
\bibitem{ARGUS:1985qda}
ARGUS collaboration, H.~Albrecht {\em et~al.},
  \ifthenelse{\boolean{articletitles}}{\emph{{Radiative decays of the~$\TwoS$
  into the~three~$\Pchi_\bquark$~states.}},
  }{}\href{https://doi.org/10.1016/0370-2693(85)91338-3}{Phys.\ Lett.\
  \textbf{B160} (1985) 331}\relax
\mciteBstWouldAddEndPuncttrue
\mciteSetBstMidEndSepPunct{\mcitedefaultmidpunct}
{\mcitedefaultendpunct}{\mcitedefaultseppunct}\relax
\EndOfBibitem
\bibitem{Heintz:1992cv}
CUSB collaboration, U.~Heintz {\em et~al.},
  \ifthenelse{\boolean{articletitles}}{\emph{{$\bquark\bquarkbar$~spectroscopy
  from the~$\ThreeS$~state}},
  }{}\href{https://doi.org/10.1103/PhysRevD.46.1928}{Phys.\ Rev.\  \textbf{D46}
  (1992) 1928}\relax
\mciteBstWouldAddEndPuncttrue
\mciteSetBstMidEndSepPunct{\mcitedefaultmidpunct}
{\mcitedefaultendpunct}{\mcitedefaultseppunct}\relax
\EndOfBibitem
\bibitem{LHCb-PAPER-2014-040}
LHCb collaboration, R.~Aaij {\em et~al.},
  \ifthenelse{\boolean{articletitles}}{\emph{{Measurement of the
  $\Pchi_{\bquark}(3\PP)$~mass and of the~relative rate of
  $\Pchi_{\bquark1}(1\PP)$ and $\Pchi_{\bquark2}(1\PP)$~production}},
  }{}\href{https://doi.org/10.1007/JHEP10(2014)088}{JHEP \textbf{10} (2014)
  088}, \href{http://arxiv.org/abs/1409.1408}{{\normalfont\ttfamily
  arXiv:1409.1408}}\relax
\mciteBstWouldAddEndPuncttrue
\mciteSetBstMidEndSepPunct{\mcitedefaultmidpunct}
{\mcitedefaultendpunct}{\mcitedefaultseppunct}\relax
\EndOfBibitem
\bibitem{LHCB-PAPER-2021-024}
LHCb collaboration, R.~Aaij {\em et~al.},
  \ifthenelse{\boolean{articletitles}}{\emph{{Measurement of the~$\PW$~boson
  mass}}, }{}\href{https://doi.org/10.1007/JHEP01(2022)036}{JHEP \textbf{01}
  (2022) 036}, \href{http://arxiv.org/abs/2109.01113}{{\normalfont\ttfamily
  arXiv:2109.01113}}\relax
\mciteBstWouldAddEndPuncttrue
\mciteSetBstMidEndSepPunct{\mcitedefaultmidpunct}
{\mcitedefaultendpunct}{\mcitedefaultseppunct}\relax
\EndOfBibitem
\bibitem{LHCb-DP-2008-001}
LHCb collaboration, A.~A. Alves~Jr.\ {\em et~al.},
  \ifthenelse{\boolean{articletitles}}{\emph{{The \lhcb detector at the LHC}},
  }{}\href{https://doi.org/10.1088/1748-0221/3/08/S08005}{JINST \textbf{3}
  (2008) S08005}\relax
\mciteBstWouldAddEndPuncttrue
\mciteSetBstMidEndSepPunct{\mcitedefaultmidpunct}
{\mcitedefaultendpunct}{\mcitedefaultseppunct}\relax
\EndOfBibitem
\bibitem{LHCb-DP-2014-002}
LHCb collaboration, R.~Aaij {\em et~al.},
  \ifthenelse{\boolean{articletitles}}{\emph{{LHCb detector performance}},
  }{}\href{https://doi.org/10.1142/S0217751X15300227}{Int.\ J.\ Mod.\ Phys.\
  \textbf{A30} (2015) 1530022},
  \href{http://arxiv.org/abs/1412.6352}{{\normalfont\ttfamily
  arXiv:1412.6352}}\relax
\mciteBstWouldAddEndPuncttrue
\mciteSetBstMidEndSepPunct{\mcitedefaultmidpunct}
{\mcitedefaultendpunct}{\mcitedefaultseppunct}\relax
\EndOfBibitem
\bibitem{LHCb-DP-2023-003}
LHCb collaboration, R.~Aaij {\em et~al.},
  \ifthenelse{\boolean{articletitles}}{\emph{{Momentum scale calibration of the
  LHCb spectrometer}},
  }{}\href{https://doi.org/10.1088/1748-0221/19/02/P02008}{JINST \textbf{19}
  (2024) P02008}, \href{http://arxiv.org/abs/2312.01772}{{\normalfont\ttfamily
  arXiv:2312.01772}}\relax
\mciteBstWouldAddEndPuncttrue
\mciteSetBstMidEndSepPunct{\mcitedefaultmidpunct}
{\mcitedefaultendpunct}{\mcitedefaultseppunct}\relax
\EndOfBibitem
\bibitem{LHCb-TDR-004}
LHCb collaboration, \ifthenelse{\boolean{articletitles}}{\emph{{LHCb muon
  system: Technical design report}}, }{}
  \href{http://cdsweb.cern.ch/search?p=CERN-LHCC-2001-010&f=reportnumber&action_search=Search&c=LHCb}
  {CERN-LHCC-2001-010}, 2001\relax
\mciteBstWouldAddEndPuncttrue
\mciteSetBstMidEndSepPunct{\mcitedefaultmidpunct}
{\mcitedefaultendpunct}{\mcitedefaultseppunct}\relax
\EndOfBibitem
\bibitem{LHCb-DP-2012-002}
A.~A. Alves~Jr.\ {\em et~al.},
  \ifthenelse{\boolean{articletitles}}{\emph{{Performance of the LHCb muon
  system}}, }{}\href{https://doi.org/10.1088/1748-0221/8/02/P02022}{JINST
  \textbf{8} (2013) P02022},
  \href{http://arxiv.org/abs/1211.1346}{{\normalfont\ttfamily
  arXiv:1211.1346}}\relax
\mciteBstWouldAddEndPuncttrue
\mciteSetBstMidEndSepPunct{\mcitedefaultmidpunct}
{\mcitedefaultendpunct}{\mcitedefaultseppunct}\relax
\EndOfBibitem
\bibitem{LHCb-DP-2012-004}
R.~Aaij {\em et~al.}, \ifthenelse{\boolean{articletitles}}{\emph{{The \lhcb
  trigger and its performance in 2011}},
  }{}\href{https://doi.org/10.1088/1748-0221/8/04/P04022}{JINST \textbf{8}
  (2013) P04022}, \href{http://arxiv.org/abs/1211.3055}{{\normalfont\ttfamily
  arXiv:1211.3055}}\relax
\mciteBstWouldAddEndPuncttrue
\mciteSetBstMidEndSepPunct{\mcitedefaultmidpunct}
{\mcitedefaultendpunct}{\mcitedefaultseppunct}\relax
\EndOfBibitem
\bibitem{LHCb-DP-2019-001}
R.~Aaij {\em et~al.}, \ifthenelse{\boolean{articletitles}}{\emph{{Design and
  performance of the LHCb trigger and full real-time reconstruction in Run 2 of
  the LHC}}, }{}\href{https://doi.org/10.1088/1748-0221/14/04/P04013}{JINST
  \textbf{14} (2019) P04013},
  \href{http://arxiv.org/abs/1812.10790}{{\normalfont\ttfamily
  arXiv:1812.10790}}\relax
\mciteBstWouldAddEndPuncttrue
\mciteSetBstMidEndSepPunct{\mcitedefaultmidpunct}
{\mcitedefaultendpunct}{\mcitedefaultseppunct}\relax
\EndOfBibitem
\bibitem{Borghi:2017hfp}
S.~Borghi, \ifthenelse{\boolean{articletitles}}{\emph{{Novel real-time
  alignment and calibration of the LHCb detector and its performance}},
  }{}\href{https://doi.org/10.1016/j.nima.2016.06.050}{Nucl.\ Instrum.\ Meth.\
  \textbf{A845} (2017) 560}\relax
\mciteBstWouldAddEndPuncttrue
\mciteSetBstMidEndSepPunct{\mcitedefaultmidpunct}
{\mcitedefaultendpunct}{\mcitedefaultseppunct}\relax
\EndOfBibitem
\bibitem{Cowan:2016tnm}
G.~A. Cowan, D.~C. Craik, and M.~D. Needham,
  \ifthenelse{\boolean{articletitles}}{\emph{{{\sc{RapidSim}}: An application
  for the~fast simulation of heavy-quark hadron decays}},
  }{}\href{https://doi.org/10.1016/j.cpc.2017.01.029}{Comput.\ Phys.\ Commun.\
  \textbf{214} (2017) 239},
  \href{http://arxiv.org/abs/1612.07489}{{\normalfont\ttfamily
  arXiv:1612.07489}}\relax
\mciteBstWouldAddEndPuncttrue
\mciteSetBstMidEndSepPunct{\mcitedefaultmidpunct}
{\mcitedefaultendpunct}{\mcitedefaultseppunct}\relax
\EndOfBibitem
\bibitem{Agostinelli:2002hh}
Geant4 collaboration, S.~Agostinelli {\em et~al.},
  \ifthenelse{\boolean{articletitles}}{\emph{{{\sc{Geant4}}: A simulation
  toolkit}}, }{}\href{https://doi.org/10.1016/S0168-9002(03)01368-8}{Nucl.\
  Instrum.\ Meth.\  \textbf{A506} (2003) 250}\relax
\mciteBstWouldAddEndPuncttrue
\mciteSetBstMidEndSepPunct{\mcitedefaultmidpunct}
{\mcitedefaultendpunct}{\mcitedefaultseppunct}\relax
\EndOfBibitem
\bibitem{Allison:2006ve}
Geant4 collaboration, J.~Allison {\em et~al.},
  \ifthenelse{\boolean{articletitles}}{\emph{{{\sc{Geant4}} developments and
  applications}}, }{}\href{https://doi.org/10.1109/TNS.2006.869826}{IEEE
  Trans.\ Nucl.\ Sci.\  \textbf{53} (2006) 270}\relax
\mciteBstWouldAddEndPuncttrue
\mciteSetBstMidEndSepPunct{\mcitedefaultmidpunct}
{\mcitedefaultendpunct}{\mcitedefaultseppunct}\relax
\EndOfBibitem
\bibitem{LHCb-PROC-2011-006}
M.~Clemencic {\em et~al.}, \ifthenelse{\boolean{articletitles}}{\emph{{The
  \lhcb simulation application, {\sc{Gauss}}: Design, evolution and
  experience}}, }{}\href{https://doi.org/10.1088/1742-6596/331/3/032023}{J.\
  Phys.\ Conf.\ Ser.\  \textbf{331} (2011) 032023}\relax
\mciteBstWouldAddEndPuncttrue
\mciteSetBstMidEndSepPunct{\mcitedefaultmidpunct}
{\mcitedefaultendpunct}{\mcitedefaultseppunct}\relax
\EndOfBibitem
\bibitem{Lange:2001uf}
D.~J. Lange, \ifthenelse{\boolean{articletitles}}{\emph{{The {\sc{EvtGen}}
  particle decay simulation package}},
  }{}\href{https://doi.org/10.1016/S0168-9002(01)00089-4}{Nucl.\ Instrum.\
  Meth.\  \textbf{A462} (2001) 152}\relax
\mciteBstWouldAddEndPuncttrue
\mciteSetBstMidEndSepPunct{\mcitedefaultmidpunct}
{\mcitedefaultendpunct}{\mcitedefaultseppunct}\relax
\EndOfBibitem
\bibitem{Golonka:2005pn}
P.~Golonka and Z.~Was,
  \ifthenelse{\boolean{articletitles}}{\emph{{{\sc{PHOTOS}} Monte Carlo: A
  precision tool for QED corrections in $\PZ$~and~$\PW$~decays}},
  }{}\href{https://doi.org/10.1140/epjc/s2005-02396-4}{Eur.\ Phys.\ J.\
  \textbf{C45} (2006) 97},
  \href{http://arxiv.org/abs/hep-ph/0506026}{{\normalfont\ttfamily
  arXiv:hep-ph/0506026}}\relax
\mciteBstWouldAddEndPuncttrue
\mciteSetBstMidEndSepPunct{\mcitedefaultmidpunct}
{\mcitedefaultendpunct}{\mcitedefaultseppunct}\relax
\EndOfBibitem
\bibitem{davidson2015photos}
N.~Davidson, T.~Przedzinski, and Z.~Was,
  \ifthenelse{\boolean{articletitles}}{\emph{{{\sc{PHOTOS}} interface in C++:
  Technical and physics documentation}},
  }{}\href{https://doi.org/https://doi.org/10.1016/j.cpc.2015.09.013}{Comp.\
  Phys.\ Comm.\  \textbf{199} (2016) 86},
  \href{http://arxiv.org/abs/1011.0937}{{\normalfont\ttfamily
  arXiv:1011.0937}}\relax
\mciteBstWouldAddEndPuncttrue
\mciteSetBstMidEndSepPunct{\mcitedefaultmidpunct}
{\mcitedefaultendpunct}{\mcitedefaultseppunct}\relax
\EndOfBibitem
\bibitem{Brown:1975dz}
L.~S. Brown and R.~N. Cahn, \ifthenelse{\boolean{articletitles}}{\emph{{Chiral
  symmetry and \mbox{$\decay{\Ppsi^{\prime}}{\Ppsi\pion\pion}$}~decay}},
  }{}\href{https://doi.org/10.1103/PhysRevLett.35.1}{Phys.\ Rev.\ Lett.\
  \textbf{35} (1975) 1}\relax
\mciteBstWouldAddEndPuncttrue
\mciteSetBstMidEndSepPunct{\mcitedefaultmidpunct}
{\mcitedefaultendpunct}{\mcitedefaultseppunct}\relax
\EndOfBibitem
\bibitem{Voloshin:2006ce}
M.~B. Voloshin, \ifthenelse{\boolean{articletitles}}{\emph{{Two-pion
  transitions in quarkonium revisited}},
  }{}\href{https://doi.org/10.1103/PhysRevD.74.054022}{Phys.\ Rev.\
  \textbf{D74} (2006) 054022},
  \href{http://arxiv.org/abs/hep-ph/0606258}{{\normalfont\ttfamily
  arXiv:hep-ph/0606258}}\relax
\mciteBstWouldAddEndPuncttrue
\mciteSetBstMidEndSepPunct{\mcitedefaultmidpunct}
{\mcitedefaultendpunct}{\mcitedefaultseppunct}\relax
\EndOfBibitem
\bibitem{Voloshin:2007dx}
M.~B. Voloshin, \ifthenelse{\boolean{articletitles}}{\emph{{Charmonium}},
  }{}\href{https://doi.org/10.1016/j.ppnp.2008.02.001}{Prog.\ Part.\ Nucl.\
  Phys.\  \textbf{61} (2008) 455},
  \href{http://arxiv.org/abs/0711.4556}{{\normalfont\ttfamily
  arXiv:0711.4556}}\relax
\mciteBstWouldAddEndPuncttrue
\mciteSetBstMidEndSepPunct{\mcitedefaultmidpunct}
{\mcitedefaultendpunct}{\mcitedefaultseppunct}\relax
\EndOfBibitem
\bibitem{Luchinsky:2017pby}
A.~V. Luchinsky, \ifthenelse{\boolean{articletitles}}{\emph{{Muon pair
  production in radiative decays of heavy quarkonia}},
  }{}\href{https://doi.org/10.1142/S0217732318500013}{Mod.\ Phys.\ Lett.\
  \textbf{A33} (2017) 1850001},
  \href{http://arxiv.org/abs/1709.02444}{{\normalfont\ttfamily
  arXiv:1709.02444}}\relax
\mciteBstWouldAddEndPuncttrue
\mciteSetBstMidEndSepPunct{\mcitedefaultmidpunct}
{\mcitedefaultendpunct}{\mcitedefaultseppunct}\relax
\EndOfBibitem
\bibitem{LHCb-PAPER-2011-036}
LHCb collaboration, R.~Aaij {\em et~al.},
  \ifthenelse{\boolean{articletitles}}{\emph{{Measurement of \Upsilonres
  production in \proton\proton collisions at \mbox{$\sqs=$7~\tev}}},
  }{}\href{https://doi.org/10.1140/epjc/s10052-012-2025-y}{Eur.\ Phys.\ J.\
  \textbf{C72} (2012) 2025},
  \href{http://arxiv.org/abs/1202.6579}{{\normalfont\ttfamily
  arXiv:1202.6579}}\relax
\mciteBstWouldAddEndPuncttrue
\mciteSetBstMidEndSepPunct{\mcitedefaultmidpunct}
{\mcitedefaultendpunct}{\mcitedefaultseppunct}\relax
\EndOfBibitem
\bibitem{LHcb-PAPER-2013-016}
LHCb collaboration, R.~Aaij {\em et~al.},
  \ifthenelse{\boolean{articletitles}}{\emph{{Production of \jpsi and
  \Upsilonres mesons in \proton\proton collisions at \mbox{$\sqs=8\tev$}}},
  }{}\href{https://doi.org/10.1007/JHEP06(2013)064}{JHEP \textbf{06} (2013)
  064}, \href{http://arxiv.org/abs/1304.6977}{{\normalfont\ttfamily
  arXiv:1304.6977}}\relax
\mciteBstWouldAddEndPuncttrue
\mciteSetBstMidEndSepPunct{\mcitedefaultmidpunct}
{\mcitedefaultendpunct}{\mcitedefaultseppunct}\relax
\EndOfBibitem
\bibitem{LHCb-PAPER-2013-066}
LHCb collaboration, R.~Aaij {\em et~al.},
  \ifthenelse{\boolean{articletitles}}{\emph{{Measurement of \Upsilonres
  production in \proton\proton collisions at $\sqs = 2.76\tev$}},
  }{}\href{https://doi.org/10.1140/epjc/s10052-014-2835-1}{Eur.\ Phys.\ J.\
  \textbf{C74} (2014) 2835},
  \href{http://arxiv.org/abs/1402.2539}{{\normalfont\ttfamily
  arXiv:1402.2539}}\relax
\mciteBstWouldAddEndPuncttrue
\mciteSetBstMidEndSepPunct{\mcitedefaultmidpunct}
{\mcitedefaultendpunct}{\mcitedefaultseppunct}\relax
\EndOfBibitem
\bibitem{LHCb-PAPER-2015-045}
LHCb collaboration, R.~Aaij {\em et~al.},
  \ifthenelse{\boolean{articletitles}}{\emph{{Forward production of \Upsilonres
  mesons in \proton\proton collisions at $\sqs=7$~and~$8\tev$}},
  }{}\href{https://doi.org/10.1007/JHEP11(2015)103}{JHEP \textbf{11} (2015)
  103}, \href{http://arxiv.org/abs/1509.02372}{{\normalfont\ttfamily
  arXiv:1509.02372}}\relax
\mciteBstWouldAddEndPuncttrue
\mciteSetBstMidEndSepPunct{\mcitedefaultmidpunct}
{\mcitedefaultendpunct}{\mcitedefaultseppunct}\relax
\EndOfBibitem
\bibitem{LHCb-PAPER-2018-002}
LHCb collaboration, R.~Aaij {\em et~al.},
  \ifthenelse{\boolean{articletitles}}{\emph{{Measurement of \Upsilonres
  production cross-section in \proton\proton~collisions at $\sqs=13\tev$}},
  }{}\href{https://doi.org/10.1007/JHEP07(2018)134}{JHEP \textbf{07} (2018)
  134}, \href{http://arxiv.org/abs/1804.09214}{{\normalfont\ttfamily
  arXiv:1804.09214}}\relax
\mciteBstWouldAddEndPuncttrue
\mciteSetBstMidEndSepPunct{\mcitedefaultmidpunct}
{\mcitedefaultendpunct}{\mcitedefaultseppunct}\relax
\EndOfBibitem
\bibitem{Breiman}
L.~Breiman, J.~H. Friedman, R.~A. Olshen, and C.~J. Stone, {\em Classification
  and regression trees}, Wadsworth international group, Belmont, California,
  USA, 1984\relax
\mciteBstWouldAddEndPuncttrue
\mciteSetBstMidEndSepPunct{\mcitedefaultmidpunct}
{\mcitedefaultendpunct}{\mcitedefaultseppunct}\relax
\EndOfBibitem
\bibitem{AdaBoost}
Y.~Freund and R.~E. Schapire, \ifthenelse{\boolean{articletitles}}{\emph{A
  decision-theoretic generalization of on-line learning and an application to
  boosting}, }{}\href{https://doi.org/10.1006/jcss.1997.1504}{J.\ Comput.\
  Syst.\ Sci.\  \textbf{55} (1997) 119}\relax
\mciteBstWouldAddEndPuncttrue
\mciteSetBstMidEndSepPunct{\mcitedefaultmidpunct}
{\mcitedefaultendpunct}{\mcitedefaultseppunct}\relax
\EndOfBibitem
\bibitem{Hocker:2007ht}
H.~Voss, A.~H{\"o}cker, J.~Stelzer, and F.~Tegenfeldt,
  \ifthenelse{\boolean{articletitles}}{\emph{{{\sc{TMVA}} -- Toolkit for
  multivariate data analysis with {\sc{ROOT}}}},
  }{}\href{https://doi.org/10.22323/1.050.0040}{PoS \textbf{ACAT} (2007)
  040}\relax
\mciteBstWouldAddEndPuncttrue
\mciteSetBstMidEndSepPunct{\mcitedefaultmidpunct}
{\mcitedefaultendpunct}{\mcitedefaultseppunct}\relax
\EndOfBibitem
\bibitem{TMVA4}
A.~H{\"o}cker {\em et~al.},
  \ifthenelse{\boolean{articletitles}}{\emph{{{\sc{TMVA\,4}} -- Toolkit for
  multivariate data analysis with {\sc{ROOT}}. Users guide.}},
  }{}\href{http://arxiv.org/abs/physics/0703039}{{\normalfont\ttfamily
  arXiv:physics/0703039}}\relax
\mciteBstWouldAddEndPuncttrue
\mciteSetBstMidEndSepPunct{\mcitedefaultmidpunct}
{\mcitedefaultendpunct}{\mcitedefaultseppunct}\relax
\EndOfBibitem
\bibitem{Pivk:2004ty}
M.~Pivk and F.~R. Le~Diberder,
  \ifthenelse{\boolean{articletitles}}{\emph{{sPlot: A statistical tool to
  unfold data distributions}},
  }{}\href{https://doi.org/10.1016/j.nima.2005.08.106}{Nucl.\ Instrum.\ Meth.\
  \textbf{A555} (2005) 356},
  \href{http://arxiv.org/abs/physics/0402083}{{\normalfont\ttfamily
  arXiv:physics/0402083}}\relax
\mciteBstWouldAddEndPuncttrue
\mciteSetBstMidEndSepPunct{\mcitedefaultmidpunct}
{\mcitedefaultendpunct}{\mcitedefaultseppunct}\relax
\EndOfBibitem
\bibitem{LHCb-DP-2013-002}
LHCb collaboration, R.~Aaij {\em et~al.},
  \ifthenelse{\boolean{articletitles}}{\emph{{Measurement of the track
  reconstruction efficiency at LHCb}},
  }{}\href{https://doi.org/10.1088/1748-0221/10/02/P02007}{JINST \textbf{10}
  (2015) P02007}, \href{http://arxiv.org/abs/1408.1251}{{\normalfont\ttfamily
  arXiv:1408.1251}}\relax
\mciteBstWouldAddEndPuncttrue
\mciteSetBstMidEndSepPunct{\mcitedefaultmidpunct}
{\mcitedefaultendpunct}{\mcitedefaultseppunct}\relax
\EndOfBibitem
\bibitem{LHCb-DP-2013-003}
R.~Arink {\em et~al.}, \ifthenelse{\boolean{articletitles}}{\emph{{Performance
  of the LHCb Outer Tracker}},
  }{}\href{https://doi.org/10.1088/1748-0221/9/01/P01002}{JINST \textbf{9}
  (2014) P01002}, \href{http://arxiv.org/abs/1311.3893}{{\normalfont\ttfamily
  arXiv:1311.3893}}\relax
\mciteBstWouldAddEndPuncttrue
\mciteSetBstMidEndSepPunct{\mcitedefaultmidpunct}
{\mcitedefaultendpunct}{\mcitedefaultseppunct}\relax
\EndOfBibitem
\bibitem{LHCb-DP-2017-001}
P.~d'Argent {\em et~al.}, \ifthenelse{\boolean{articletitles}}{\emph{{Improved
  performance of the LHCb Outer Tracker in LHC Run\,2}},
  }{}\href{https://doi.org/10.1088/1748-0221/12/11/P11016}{JINST \textbf{12}
  (2017) P11016}, \href{http://arxiv.org/abs/1708.00819}{{\normalfont\ttfamily
  arXiv:1708.00819}}\relax
\mciteBstWouldAddEndPuncttrue
\mciteSetBstMidEndSepPunct{\mcitedefaultmidpunct}
{\mcitedefaultendpunct}{\mcitedefaultseppunct}\relax
\EndOfBibitem
\bibitem{DeCian:2255039}
M.~De~Cian, S.~Farry, P.~Seyfert, and S.~Stahl,
  \ifthenelse{\boolean{articletitles}}{\emph{{Fast neural-net based fake track
  rejection in the LHCb reconstruction}}, }{}
  \href{http://cdsweb.cern.ch/search?p=LHCb-PUB-2017-011&f=reportnumber&action_search=Search&c=LHCb+Notes}
  {LHCb-PUB-2017-011}, 2017\relax
\mciteBstWouldAddEndPuncttrue
\mciteSetBstMidEndSepPunct{\mcitedefaultmidpunct}
{\mcitedefaultendpunct}{\mcitedefaultseppunct}\relax
\EndOfBibitem
\bibitem{LHCb-PROC-2011-008}
A.~Powell {\em et~al.}, \ifthenelse{\boolean{articletitles}}{\emph{{Particle
  identification at LHCb}}, }{}PoS \textbf{ICHEP2010} (2010) 020,
  \href{https://cdsweb.cern.ch/record/1322666?ln=en}{LHCb-PROC-2011-008}\relax
\mciteBstWouldAddEndPuncttrue
\mciteSetBstMidEndSepPunct{\mcitedefaultmidpunct}
{\mcitedefaultendpunct}{\mcitedefaultseppunct}\relax
\EndOfBibitem
\bibitem{Skwarnicki:1986xj}
T.~Skwarnicki, {\em {A study of the~radiative cascade transitions between
  the~$\PUpsilon^\prime$~and $\PUpsilon$~resonances}}, PhD thesis, Institute of
  Nuclear Physics, Krakow, 1986,
  {\href{http://inspirehep.net/record/230779/}{DESY-F31-86-02}}\relax
\mciteBstWouldAddEndPuncttrue
\mciteSetBstMidEndSepPunct{\mcitedefaultmidpunct}
{\mcitedefaultendpunct}{\mcitedefaultseppunct}\relax
\EndOfBibitem
\bibitem{LHCb-PAPER-2011-013}
LHCb collaboration, R.~Aaij {\em et~al.},
  \ifthenelse{\boolean{articletitles}}{\emph{{Observation of \jpsi-pair
  production in \proton\proton collisions at \mbox{$\sqs=7\tev$}}},
  }{}\href{https://doi.org/10.1016/j.physletb.2011.12.015}{Phys.\ Lett.\
  \textbf{B707} (2012) 52},
  \href{http://arxiv.org/abs/1109.0963}{{\normalfont\ttfamily
  arXiv:1109.0963}}\relax
\mciteBstWouldAddEndPuncttrue
\mciteSetBstMidEndSepPunct{\mcitedefaultmidpunct}
{\mcitedefaultendpunct}{\mcitedefaultseppunct}\relax
\EndOfBibitem
\bibitem{LHcb-PAPER-2013-011}
LHCb collaboration, R.~Aaij {\em et~al.},
  \ifthenelse{\boolean{articletitles}}{\emph{{Precision measurement of \D meson
  mass differences}}, }{}\href{https://doi.org/10.1007/JHEP06(2013)065}{JHEP
  \textbf{06} (2013) 065},
  \href{http://arxiv.org/abs/1304.6865}{{\normalfont\ttfamily
  arXiv:1304.6865}}\relax
\mciteBstWouldAddEndPuncttrue
\mciteSetBstMidEndSepPunct{\mcitedefaultmidpunct}
{\mcitedefaultendpunct}{\mcitedefaultseppunct}\relax
\EndOfBibitem
\bibitem{LHCb-PAPER-2011-035}
LHCb collaboration, R.~Aaij {\em et~al.},
  \ifthenelse{\boolean{articletitles}}{\emph{{Measurement of \bquark-hadron
  masses}}, }{}\href{https://doi.org/10.1016/j.physletb.2012.01.058}{Phys.\
  Lett.\  \textbf{B708} (2012) 241},
  \href{http://arxiv.org/abs/1112.4896}{{\normalfont\ttfamily
  arXiv:1112.4896}}\relax
\mciteBstWouldAddEndPuncttrue
\mciteSetBstMidEndSepPunct{\mcitedefaultmidpunct}
{\mcitedefaultendpunct}{\mcitedefaultseppunct}\relax
\EndOfBibitem
\bibitem{LHCb-PAPER-2014-015}
LHCb collaboration, R.~Aaij {\em et~al.},
  \ifthenelse{\boolean{articletitles}}{\emph{{Study of \Upsilonres production
  and cold nuclear matter effects in $\proton\mathrm{Pb}$ collisions at $\sqsnn
  = 5\tev$}}, }{}\href{https://doi.org/10.1007/JHEP07(2014)094}{JHEP
  \textbf{07} (2014) 094},
  \href{http://arxiv.org/abs/1405.5152}{{\normalfont\ttfamily
  arXiv:1405.5152}}\relax
\mciteBstWouldAddEndPuncttrue
\mciteSetBstMidEndSepPunct{\mcitedefaultmidpunct}
{\mcitedefaultendpunct}{\mcitedefaultseppunct}\relax
\EndOfBibitem
\bibitem{LHCb-PAPER-2015-046}
LHCb collaboration, R.~Aaij {\em et~al.},
  \ifthenelse{\boolean{articletitles}}{\emph{{Production of associated
  \Upsilonres and open charm hadrons in \proton\proton collisions at $\sqs =
  7$~and~$8\tev$ via double parton scattering}},
  }{}\href{https://doi.org/10.1007/JHEP07(2016)052}{JHEP \textbf{07} (2016)
  052}, \href{http://arxiv.org/abs/1510.05949}{{\normalfont\ttfamily
  arXiv:1510.05949}}\relax
\mciteBstWouldAddEndPuncttrue
\mciteSetBstMidEndSepPunct{\mcitedefaultmidpunct}
{\mcitedefaultendpunct}{\mcitedefaultseppunct}\relax
\EndOfBibitem
\bibitem{LHCb-PAPER-2017-028}
LHCb collaboration, R.~Aaij {\em et~al.},
  \ifthenelse{\boolean{articletitles}}{\emph{{Measurement of the
  $\Upsilonres(\Pn\PS)$ polarizations in \proton\proton collisions at $\sqs=7$
  and 8\tev}}, }{}\href{https://doi.org/10.1007/JHEP12(2017)110}{JHEP
  \textbf{12} (2017) 110},
  \href{http://arxiv.org/abs/1709.01301}{{\normalfont\ttfamily
  arXiv:1709.01301}}\relax
\mciteBstWouldAddEndPuncttrue
\mciteSetBstMidEndSepPunct{\mcitedefaultmidpunct}
{\mcitedefaultendpunct}{\mcitedefaultseppunct}\relax
\EndOfBibitem
\bibitem{LHCb-DP-2012-003}
M.~Adinolfi {\em et~al.},
  \ifthenelse{\boolean{articletitles}}{\emph{{Performance of the \lhcb RICH
  detector at the LHC}},
  }{}\href{https://doi.org/10.1140/epjc/s10052-013-2431-9}{Eur.\ Phys.\ J.\
  \textbf{C73} (2013) 2431},
  \href{http://arxiv.org/abs/1211.6759}{{\normalfont\ttfamily
  arXiv:1211.6759}}\relax
\mciteBstWouldAddEndPuncttrue
\mciteSetBstMidEndSepPunct{\mcitedefaultmidpunct}
{\mcitedefaultendpunct}{\mcitedefaultseppunct}\relax
\EndOfBibitem
\bibitem{PDG2024}
Particle Data Group, S.~Navas {\em et~al.},
  \ifthenelse{\boolean{articletitles}}{\emph{{\href{http://pdg.lbl.gov/2024}{Review
  of particle physics}}},
  }{}\href{https://doi.org/10.1103/PhysRevD.110.030001}{Phys.\ Rev.\
  \textbf{D110} (2024) 030001}\relax
\mciteBstWouldAddEndPuncttrue
\mciteSetBstMidEndSepPunct{\mcitedefaultmidpunct}
{\mcitedefaultendpunct}{\mcitedefaultseppunct}\relax
\EndOfBibitem
\bibitem{Hulsbergen:2005pu}
W.~D. Hulsbergen, \ifthenelse{\boolean{articletitles}}{\emph{{Decay chain
  fitting with a Kalman filter}},
  }{}\href{https://doi.org/10.1016/j.nima.2005.06.078}{Nucl.\ Instrum.\ Meth.\
  \textbf{A552} (2005) 566},
  \href{http://arxiv.org/abs/physics/0503191}{{\normalfont\ttfamily
  arXiv:physics/0503191}}\relax
\mciteBstWouldAddEndPuncttrue
\mciteSetBstMidEndSepPunct{\mcitedefaultmidpunct}
{\mcitedefaultendpunct}{\mcitedefaultseppunct}\relax
\EndOfBibitem
\bibitem{LHCb-PAPER-2020-008}
LHCb collaboration, R.~Aaij {\em et~al.},
  \ifthenelse{\boolean{articletitles}}{\emph{{Study of the line shape of the
  $\chicone(3872)$ state}},
  }{}\href{https://doi.org/10.1103/PhysRevD.102.092005}{Phys.\ Rev.\
  \textbf{D102} (2020) 092005},
  \href{http://arxiv.org/abs/2005.13419}{{\normalfont\ttfamily
  arXiv:2005.13419}}\relax
\mciteBstWouldAddEndPuncttrue
\mciteSetBstMidEndSepPunct{\mcitedefaultmidpunct}
{\mcitedefaultendpunct}{\mcitedefaultseppunct}\relax
\EndOfBibitem
\bibitem{LHCb-PAPER-2020-009}
LHCb collaboration, R.~Aaij {\em et~al.},
  \ifthenelse{\boolean{articletitles}}{\emph{{Study of the $\psires_2(3823)$
  and $\chicone(3872)$ states in
  \mbox{$\decay{\Bu}{\left(\jpsi\pip\pim\right)\Kp}$}~decays}},
  }{}\href{https://doi.org/10.1007/JHEP08(2020)123}{JHEP \textbf{08} (2020)
  123}, \href{http://arxiv.org/abs/2005.13422}{{\normalfont\ttfamily
  arXiv:2005.13422}}\relax
\mciteBstWouldAddEndPuncttrue
\mciteSetBstMidEndSepPunct{\mcitedefaultmidpunct}
{\mcitedefaultendpunct}{\mcitedefaultseppunct}\relax
\EndOfBibitem
\bibitem{LHCb-PAPER-2020-035}
LHCb collaboration, R.~Aaij {\em et~al.},
  \ifthenelse{\boolean{articletitles}}{\emph{{Study of \mbox{$\decay{\Bs}{\jpsi
  \pip \pim \Kp \Km}$}~decays}},
  }{}\href{https://doi.org/10.1007/JHEP02(2021)024}{JHEP \textbf{02} (2021)
  024}, \href{http://arxiv.org/abs/2011.01867}{{\normalfont\ttfamily
  arXiv:2011.01867}}\relax
\mciteBstWouldAddEndPuncttrue
\mciteSetBstMidEndSepPunct{\mcitedefaultmidpunct}
{\mcitedefaultendpunct}{\mcitedefaultseppunct}\relax
\EndOfBibitem
\bibitem{LHCb-PAPER-2021-031}
LHCb collaboration, R.~Aaij {\em et~al.},
  \ifthenelse{\boolean{articletitles}}{\emph{{Observation of an exotic narrow
  doubly charmed tetraquark}},
  }{}\href{https://doi.org/10.1038/s41567-022-01614-y}{Nature Physics
  \textbf{18} (2022) 751},
  \href{http://arxiv.org/abs/2109.01038}{{\normalfont\ttfamily
  arXiv:2109.01038}}\relax
\mciteBstWouldAddEndPuncttrue
\mciteSetBstMidEndSepPunct{\mcitedefaultmidpunct}
{\mcitedefaultendpunct}{\mcitedefaultseppunct}\relax
\EndOfBibitem
\bibitem{LHCb-PAPER-2021-032}
LHCb collaboration, R.~Aaij {\em et~al.},
  \ifthenelse{\boolean{articletitles}}{\emph{{Study of the doubly charmed
  tetraquark $\PT^+_{\cquark\cquark}$}},
  }{}\href{https://doi.org/10.1038/s41467-022-30206-w}{Nature Communications
  \textbf{13} (2022) 3351},
  \href{http://arxiv.org/abs/2109.01056}{{\normalfont\ttfamily
  arXiv:2109.01056}}\relax
\mciteBstWouldAddEndPuncttrue
\mciteSetBstMidEndSepPunct{\mcitedefaultmidpunct}
{\mcitedefaultendpunct}{\mcitedefaultseppunct}\relax
\EndOfBibitem
\bibitem{ARGUS:1986gsf}
ARGUS collaboration, H.~Albrecht {\em et~al.},
  \ifthenelse{\boolean{articletitles}}{\emph{{The~hadronic transitions from
  \TwoS~to~\OneS}}, }{}\href{https://doi.org/10.1007/BF01570762}{Z.\ Phys.\
  \textbf{C35} (1987) 283}\relax
\mciteBstWouldAddEndPuncttrue
\mciteSetBstMidEndSepPunct{\mcitedefaultmidpunct}
{\mcitedefaultendpunct}{\mcitedefaultseppunct}\relax
\EndOfBibitem
\bibitem{CLEO:1993fsd}
CLEO collaboration, F.~Butler {\em et~al.},
  \ifthenelse{\boolean{articletitles}}{\emph{{Analysis of hadronic transitions
  in \ThreeS~decays}}, }{}\href{https://doi.org/10.1103/PhysRevD.49.40}{Phys.\
  Rev.\  \textbf{D49} (1994) 40}\relax
\mciteBstWouldAddEndPuncttrue
\mciteSetBstMidEndSepPunct{\mcitedefaultmidpunct}
{\mcitedefaultendpunct}{\mcitedefaultseppunct}\relax
\EndOfBibitem
\bibitem{Segovia:2016xqb}
J.~Segovia, P.~G. Ortega, D.~R. Entem, and F.~Fern\'andez,
  \ifthenelse{\boolean{articletitles}}{\emph{{Bottomonium spectrum revisited}},
  }{}\href{https://doi.org/10.1103/PhysRevD.93.074027}{Phys.\ Rev.\
  \textbf{D93} (2016) 074027},
  \href{http://arxiv.org/abs/1601.05093}{{\normalfont\ttfamily
  arXiv:1601.05093}}\relax
\mciteBstWouldAddEndPuncttrue
\mciteSetBstMidEndSepPunct{\mcitedefaultmidpunct}
{\mcitedefaultendpunct}{\mcitedefaultseppunct}\relax
\EndOfBibitem
\bibitem{Deng:2016ktl}
W.-J. Deng, H.~Liu, L.-C. Gui, and X.-H. Zhong,
  \ifthenelse{\boolean{articletitles}}{\emph{{Spectrum and electromagnetic
  transitions of bottomonium}},
  }{}\href{https://doi.org/10.1103/PhysRevD.95.074002}{Phys.\ Rev.\
  \textbf{D95} (2017) 074002},
  \href{http://arxiv.org/abs/1607.04696}{{\normalfont\ttfamily
  arXiv:1607.04696}}\relax
\mciteBstWouldAddEndPuncttrue
\mciteSetBstMidEndSepPunct{\mcitedefaultmidpunct}
{\mcitedefaultendpunct}{\mcitedefaultseppunct}\relax
\EndOfBibitem
\bibitem{Asghar:2023fvk}
I.~Asghar and N.~Akbar, \ifthenelse{\boolean{articletitles}}{\emph{{Spectrum
  and decay properties of bottomonium mesons}},
  }{}\href{https://doi.org/10.1140/epja/s10050-024-01279-6}{Eur.\ Phys.\ J.\
  \textbf{A60} (2024) 58},
  \href{http://arxiv.org/abs/2309.15438}{{\normalfont\ttfamily
  arXiv:2309.15438}}\relax
\mciteBstWouldAddEndPuncttrue
\mciteSetBstMidEndSepPunct{\mcitedefaultmidpunct}
{\mcitedefaultendpunct}{\mcitedefaultseppunct}\relax
\EndOfBibitem
\bibitem{Wilks:1938dza}
S.~S. Wilks, \ifthenelse{\boolean{articletitles}}{\emph{{The large-sample
  distribution of the likelihood ratio for testing composite hypotheses}},
  }{}\href{https://doi.org/10.1214/aoms/1177732360}{Ann.\ Math.\ Stat.\
  \textbf{9} (1938) 60}\relax
\mciteBstWouldAddEndPuncttrue
\mciteSetBstMidEndSepPunct{\mcitedefaultmidpunct}
{\mcitedefaultendpunct}{\mcitedefaultseppunct}\relax
\EndOfBibitem
\bibitem{Byckling}
E.~Byckling and K.~Kajantie, {\em Particle kinematics}, John Wiley \& Sons
  Inc., New York, 1973\relax
\mciteBstWouldAddEndPuncttrue
\mciteSetBstMidEndSepPunct{\mcitedefaultmidpunct}
{\mcitedefaultendpunct}{\mcitedefaultseppunct}\relax
\EndOfBibitem
\bibitem{PDG2022}
Particle Data Group, R.~L. Workman {\em et~al.},
  \ifthenelse{\boolean{articletitles}}{\emph{{\href{http://pdg.lbl.gov/2022/}{Review
  of particle physics}}}, }{}\href{https://doi.org/10.1093/ptep/ptac097}{Prog.\
  Theor.\ Exp.\ Phys.\  \textbf{2022} (2022) 083C01}\relax
\mciteBstWouldAddEndPuncttrue
\mciteSetBstMidEndSepPunct{\mcitedefaultmidpunct}
{\mcitedefaultendpunct}{\mcitedefaultseppunct}\relax
\EndOfBibitem
\bibitem{BaBar:2011god}
BaBar collaboration, J.~P. Lees {\em et~al.},
  \ifthenelse{\boolean{articletitles}}{\emph{{Study of
  $\decay{\PUpsilon(3\PS,2\PS)}{\Peta \PUpsilon(1\PS)}$ and
  $\decay{\PUpsilon(3\PS,2\PS)}{\pip\pim \PUpsilon(1\PS)}$~hadronic
  trasitions}}, }{}\href{https://doi.org/10.1103/PhysRevD.84.092003}{Phys.\
  Rev.\  \textbf{D84} (2011) 092003},
  \href{http://arxiv.org/abs/1108.5874}{{\normalfont\ttfamily
  arXiv:1108.5874}}\relax
\mciteBstWouldAddEndPuncttrue
\mciteSetBstMidEndSepPunct{\mcitedefaultmidpunct}
{\mcitedefaultendpunct}{\mcitedefaultseppunct}\relax
\EndOfBibitem
\bibitem{Artamonov:1983vz}
MD-1 collaboration, A.~S. Artamonov {\em et~al.},
  \ifthenelse{\boolean{articletitles}}{\emph{{A~high precision measurement of
  the~$\PUpsilon$, $\PUpsilon^\prime$ and $\PUpsilon^{\prime\prime}$~meson
  masses}}, }{}\href{https://doi.org/10.1016/0370-2693(84)90243-0}{Phys.\
  Lett.\  \textbf{B137} (1984) 272}\relax
\mciteBstWouldAddEndPuncttrue
\mciteSetBstMidEndSepPunct{\mcitedefaultmidpunct}
{\mcitedefaultendpunct}{\mcitedefaultseppunct}\relax
\EndOfBibitem
\bibitem{Baru:1986mi}
MD-1 collaboration, S.~E. Baru {\em et~al.},
  \ifthenelse{\boolean{articletitles}}{\emph{{New measurement of
  the~$\PUpsilon$~meson mass}},
  }{}\href{https://doi.org/10.1007/BF01571803}{Z.\ ~Phys.\  \textbf{C30} (1986)
  551}, Erratum \href{https://doi.org/10.1007/BF01550789}{ibid.\   \textbf{C32}
  (1986) 622}\relax
\mciteBstWouldAddEndPuncttrue
\mciteSetBstMidEndSepPunct{\mcitedefaultmidpunct}
{\mcitedefaultendpunct}{\mcitedefaultseppunct}\relax
\EndOfBibitem
\bibitem{Baru:1992jnf}
MD-1 collaboration, S.~E. Baru {\em et~al.},
  \ifthenelse{\boolean{articletitles}}{\emph{{Determination of
  the~$\OneS$~leptonic width}},
  }{}\href{https://doi.org/10.1007/BF01474726}{Z.\ Phys.\  \textbf{C56} (1992)
  547}\relax
\mciteBstWouldAddEndPuncttrue
\mciteSetBstMidEndSepPunct{\mcitedefaultmidpunct}
{\mcitedefaultendpunct}{\mcitedefaultseppunct}\relax
\EndOfBibitem
\bibitem{MacKay:1984kv}
{W.\ ~W.\ ~MacKay {\em{et.\ al.\ }} and CUSB collaboration},
  \ifthenelse{\boolean{articletitles}}{\emph{{Measurement of the
  $\PUpsilon$~mass}}, }{}\href{https://doi.org/10.1103/PhysRevD.29.2483}{Phys.\
  Rev.\  \textbf{D29} (1984) 2483}\relax
\mciteBstWouldAddEndPuncttrue
\mciteSetBstMidEndSepPunct{\mcitedefaultmidpunct}
{\mcitedefaultendpunct}{\mcitedefaultseppunct}\relax
\EndOfBibitem
\bibitem{ARGUS:1983mhk}
{D.\ ~P.\ ~Barber {\em{et al.\ }} and ARGUS and Crystal Ball collaborations},
  \ifthenelse{\boolean{articletitles}}{\emph{{A~precision measurement of the
  $\PUpsilon^\prime$~meson mass}},
  }{}\href{https://doi.org/10.1016/0370-2693(84)90323-X}{Phys.\ Lett.\
  \textbf{B135} (1984) 498}\relax
\mciteBstWouldAddEndPuncttrue
\mciteSetBstMidEndSepPunct{\mcitedefaultmidpunct}
{\mcitedefaultendpunct}{\mcitedefaultseppunct}\relax
\EndOfBibitem
\bibitem{Shamov:2022ajx}
A.~G. Shamov and O.~L. Rezanova,
  \ifthenelse{\boolean{articletitles}}{\emph{{Revision of results on
  $\PUpsilon(1\PS)$, $\PUpsilon(2\PS)$, and $\PUpsilon(3\PS)$~masses}},
  }{}\href{https://doi.org/10.1016/j.physletb.2023.137766}{Phys.\ Lett.\
  \textbf{B839} (2023) 137766},
  \href{http://arxiv.org/abs/2210.13930}{{\normalfont\ttfamily
  arXiv:2210.13930}}\relax
\mciteBstWouldAddEndPuncttrue
\mciteSetBstMidEndSepPunct{\mcitedefaultmidpunct}
{\mcitedefaultendpunct}{\mcitedefaultseppunct}\relax
\EndOfBibitem
\bibitem{Kuraev:1985hb}
E.~A. Kuraev and V.~S. Fadin,
  \ifthenelse{\boolean{articletitles}}{\emph{{On~radiative corrections to
  \epem~single photon annihilation at high\nobreakdash-energy}}, }{}Sov.\ J.\
  Nucl.\ Phys.\  \textbf{41} (1985) 466\relax
\mciteBstWouldAddEndPuncttrue
\mciteSetBstMidEndSepPunct{\mcitedefaultmidpunct}
{\mcitedefaultendpunct}{\mcitedefaultseppunct}\relax
\EndOfBibitem
\bibitem{Azimov:1975ft}
Y.~I. Azimov, A.~I. Vainshtein, L.~N. Lipatov, and V.~A. Khoze,
  \ifthenelse{\boolean{articletitles}}{\emph{{Electromagnetic corrections to
  the production of narrow resonances in colliding \epem~beams}}, }{}Pisma Zh.\
  Eksp.\ Teor.\ Fiz.\  \textbf{21} (1975) 378\relax
\mciteBstWouldAddEndPuncttrue
\mciteSetBstMidEndSepPunct{\mcitedefaultmidpunct}
{\mcitedefaultendpunct}{\mcitedefaultseppunct}\relax
\EndOfBibitem
\bibitem{Anashin:2011ku}
KEDR collaboration, V.~V. Anashin {\em et~al.},
  \ifthenelse{\boolean{articletitles}}{\emph{{Measurement of main parameters of
  the~\psitwos~resonance}},
  }{}\href{https://doi.org/10.1016/j.physletb.2012.04.009}{Phys.\ Lett.\
  \textbf{B711} (2012) 280},
  \href{http://arxiv.org/abs/1109.4215}{{\normalfont\ttfamily
  arXiv:1109.4215}}\relax
\mciteBstWouldAddEndPuncttrue
\mciteSetBstMidEndSepPunct{\mcitedefaultmidpunct}
{\mcitedefaultendpunct}{\mcitedefaultseppunct}\relax
\EndOfBibitem
\bibitem{Cohen:1987fr}
E.~R. Cohen and B.~N. Taylor, \ifthenelse{\boolean{articletitles}}{\emph{{The
  1986 adjustment of the fundamental physical constants}},
  }{}\href{https://doi.org/10.1103/RevModPhys.59.1121}{Rev.\ Mod.\ Phys.\
  \textbf{59} (1987) 1121}\relax
\mciteBstWouldAddEndPuncttrue
\mciteSetBstMidEndSepPunct{\mcitedefaultmidpunct}
{\mcitedefaultendpunct}{\mcitedefaultseppunct}\relax
\EndOfBibitem
\bibitem{BaBar:2011krt}
BaBar collaboration, J.~P. Lees {\em et~al.},
  \ifthenelse{\boolean{articletitles}}{\emph{{Study of di-pion bottomonium
  transitions and search for the $\Ph_\bquark(1\PP)$~state}},
  }{}\href{https://doi.org/10.1103/PhysRevD.84.011104}{Phys.\ Rev.\
  \textbf{D84} (2011) 011104},
  \href{http://arxiv.org/abs/1105.4234}{{\normalfont\ttfamily
  arXiv:1105.4234}}\relax
\mciteBstWouldAddEndPuncttrue
\mciteSetBstMidEndSepPunct{\mcitedefaultmidpunct}
{\mcitedefaultendpunct}{\mcitedefaultseppunct}\relax
\EndOfBibitem
\bibitem{LHCb-PAPER-2018-027}
LHCb collaboration, R.~Aaij {\em et~al.},
  \ifthenelse{\boolean{articletitles}}{\emph{{Search for beautiful tetraquarks
  in the $\OneS\mumu$ invariant-mass spectrum}},
  }{}\href{https://doi.org/10.1007/JHEP10(2018)086}{JHEP \textbf{10} (2018)
  086}, \href{http://arxiv.org/abs/1806.09707}{{\normalfont\ttfamily
  arXiv:1806.09707}}\relax
\mciteBstWouldAddEndPuncttrue
\mciteSetBstMidEndSepPunct{\mcitedefaultmidpunct}
{\mcitedefaultendpunct}{\mcitedefaultseppunct}\relax
\EndOfBibitem
\bibitem{LHCb-DP-2022-002}
LHCb collaboration, R.~Aaij {\em et~al.},
  \ifthenelse{\boolean{articletitles}}{\emph{{The LHCb Upgrade~I}},
  }{}\href{https://doi.org/10.1088/1748-0221/19/05/P05065}{JINST \textbf{19}
  (2024) P05065}, \href{http://arxiv.org/abs/2305.10515}{{\normalfont\ttfamily
  arXiv:2305.10515}}\relax
\mciteBstWouldAddEndPuncttrue
\mciteSetBstMidEndSepPunct{\mcitedefaultmidpunct}
{\mcitedefaultendpunct}{\mcitedefaultseppunct}\relax
\EndOfBibitem
\bibitem{LHCb-TDR-023}
LHCb collaboration, \ifthenelse{\boolean{articletitles}}{\emph{{LHCb Framework
  TDR for the LHCb Upgrade II: Opportunities in flavour physics, and beyond, in
  the HL$-$LHC era}}, }{}
  \href{http://cdsweb.cern.ch/search?p=CERN-LHCC-2021-012&f=reportnumber&action_search=Search&c=LHCb}
  {CERN-LHCC-2021-012}, 2022\relax
\mciteBstWouldAddEndPuncttrue
\mciteSetBstMidEndSepPunct{\mcitedefaultmidpunct}
{\mcitedefaultendpunct}{\mcitedefaultseppunct}\relax
\EndOfBibitem
\end{mcitethebibliography}

\newpage
% LHCb collaboration author list
% Data extracted on August 8th, 2024 at 11:36am for paper reference LHCb-PAPER-2024-025
\centerline
{\large\bf LHCb collaboration}
\begin
{flushleft}
\small
R.~Aaij$^{37}$\lhcborcid{0000-0003-0533-1952},
A.S.W.~Abdelmotteleb$^{56}$\lhcborcid{0000-0001-7905-0542},
C.~Abellan~Beteta$^{50}$,
F.~Abudin{\'e}n$^{56}$\lhcborcid{0000-0002-6737-3528},
T.~Ackernley$^{60}$\lhcborcid{0000-0002-5951-3498},
A. A. ~Adefisoye$^{68}$\lhcborcid{0000-0003-2448-1550},
B.~Adeva$^{46}$\lhcborcid{0000-0001-9756-3712},
M.~Adinolfi$^{54}$\lhcborcid{0000-0002-1326-1264},
P.~Adlarson$^{81}$\lhcborcid{0000-0001-6280-3851},
C.~Agapopoulou$^{14}$\lhcborcid{0000-0002-2368-0147}
% ...

{\footnotesize \it

% ...
$^{14}$Universit{\'e} Paris-Saclay, CNRS/IN2P3, IJCLab, Orsay, France\\
% ...
$^{37}$Nikhef National Institute for Subatomic Physics, Amsterdam, Netherlands\\
% ...
$^{46}$Instituto Galego de F{\'\i}sica de Altas Enerx{\'\i}as (IGFAE), Universidade de Santiago de Compostela, Santiago de Compostela, Spain\\
% ...
$^{50}$Physik-Institut, Universit{\"a}t Z{\"u}rich, Z{\"u}rich, Switzerland\\
% ...
$^{54}$H.H. Wills Physics Laboratory, University of Bristol, Bristol, United Kingdom\\
% ...
$^{56}$Department of Physics, University of Warwick, Coventry, United Kingdom\\
% ...
$^{59}$School of Physics and Astronomy, University of Glasgow, Glasgow, United Kingdom\\
$^{60}$Oliver Lodge Laboratory, University of Liverpool, Liverpool, United Kingdom\\
% ...
$^{68}$Syracuse University, Syracuse, NY, United States\\
% ...
$^{81}$Department of Physics and Astronomy, Uppsala University, Uppsala, Sweden, associated to $^{59}$\\
% ...
}
\end{flushleft}

\end{document}